\documentclass[fleqn,usenatbib]{mnras}

\usepackage{newtxtext,newtxmath}

\usepackage[T1]{fontenc}

\DeclareRobustCommand{\VAN}[3]{#2}
\let\VANthebibliography\thebibliography
\def\thebibliography{\DeclareRobustCommand{\VAN}[3]{##3}\VANthebibliography}

\usepackage{graphicx}	
\usepackage{amsmath}	
\usepackage{hyperref}
\usepackage{longtable}
\usepackage{tabularx}
\usepackage{multirow}
\usepackage{rotating}
\usepackage{lscape}     
\usepackage[dvipsnames]{xcolor}
\usepackage[caption = false]{subfig}
\usepackage{float}
\usepackage{soul}
\usepackage[normalem]{ulem}
\usepackage{animate}    
\usepackage{booktabs}

\usepackage{graphicx} 
\usepackage{url}
\usepackage{amsmath}
\usepackage{orcidlink}

\title[EM Predictions for Neutron Star Mergers in LVK O4 and O5]{Predictions for Electromagnetic Counterparts to Neutron Star Mergers Discovered during LIGO-Virgo-KAGRA Observing Runs 4 and 5}

\author[Shah et al.]{
Ved~G.~Shah$^{1,2}$
\orcidlink{0009-0009-1590-2318},
\thanks{E-mail: vedgs2@illinois.edu}
Gautham~Narayan$^{1, 4, 5}$
\orcidlink{0000-0001-6022-0484},
Haille~M.~L.~Perkins$^{1,4,5}$
\orcidlink{0009-0000-5561-9116},
Ryan~J.~Foley$^{6}$ 
\orcidlink{0000-0002-2445-5275},
\newauthor Deep~Chatterjee$^{7}$ 
\orcidlink{0000-0003-0038-5468},
Bryce~Cousins$^{3,5}$ 
\orcidlink{0000-0002-7026-1340},
Phillip Macias$^{6}$
\orcidlink{0000-0002-9946-4635}\\
$^{1}$ Department of Astronomy, University of Illinois at Urbana-Champaign, 1002 West Green Street, Urbana, IL 61801, USA \\
$^{2}$ Department of Computer Science, University of Illinois at Urbana-Champaign,  Urbana, IL 61801, USA \\
$^{3}$ Department of Physics, University of Illinois Urbana-Champaign, Urbana, IL 61801, USA \\
$^{4}$ Center for AstroPhysical Surveys, National Center for Supercomputing Applications, Urbana, IL, 61801, USA \\
$^{5}$ Illinois Center for Advanced Studies of the Universe, University of Illinois Urbana-Champaign, Urbana, IL 61801, USA \\
$^{6}$ Department of Astronomy and Astrophysics, University of California Santa Cruz, Santa Cruz, CA 95064, USA \\
$^{7}$ LIGO Laboratory and Kavli Institute for Astrophysics and Space Research, Massachusetts Institute of Technology, 185 Albany Street, Cambridge,\\ Massachusetts 02139, USA \\
}

\date{Accepted XXX. Received YYY; in original form ZZZ}

\begin{document}

\maketitle

\begin{abstract}
We present a comprehensive, configurable open-source software framework for estimating the rate of electromagnetic detection of kilonovae (KNe) associated with gravitational wave detections of binary neutron star (BNS) mergers.  We simulate the current LIGO-Virgo-KAGRA (LVK) observing run (O4) using current sensitivity and uptime values as well as using predicted sensitivites for the next observing run (O5). We find the number of discoverable kilonovae during LVK O4 to be ${ 1}_{- 1}^{+ 4}$ or ${ 2 }_{- 2 }^{+ 3 }$, (at 90\% confidence) depending on the distribution of NS masses in coalescing binaries, with the number increasing by an order of magnitude during O5 to ${ 19 }_{- 11 }^{+ 24 }$. Regardless of mass model, we predict at most five detectable KNe (at 95-per cent confidence) in O4.  We also produce optical and near-infrared light curves that correspond to the physical properties of each merging system.  We have collated important information for allocating observing resources for search and follow-up observations, including distributions of peak magnitudes in several broad bands and timescales for which specific facilities can detect each KN. The framework is easily adaptable, and new simulations can quickly be produced in response to updated information such as refined merger rates and NS mass distributions.  Finally, we compare our suite of simulations to the thus-far completed portion of O4 (as of October 14, 2023), finding a median number of discoverable KNe of 0 and a 95-percentile upper limit of 2, consistent with no detections so far in O4.
\end{abstract}


\section{Introduction}

Currently, observable gravitational waves are primarily produced by the coalescence of binary compact objects \citep{LIGOScientific:2016aoc,2019PhRvX...9c1040A, 2021ApJ...913L...7A, 2021arXiv211103606T}. Specifically, binary neutron star (BNS) mergers, like GW170817~\citep{LIGOScientific:2017vwq}, are of interest as these events can yield a post-merger, electromagnetic counterpart known as a kilonova~\citep{LIGOScientific:2017ync}. These transient events are fueled by the radioactive decay of heavy nuclei which are synthesized through r-process nucleosynthesis reactions possible given the neutron rich environment. Under certain conditions, black hole-neutron star mergers can produce kilonovae as well; however, it is much less likely \citep{Fragione_2021}, so we focus on BNS mergers here.

\begin{figure*}

    \includegraphics[width=\linewidth]{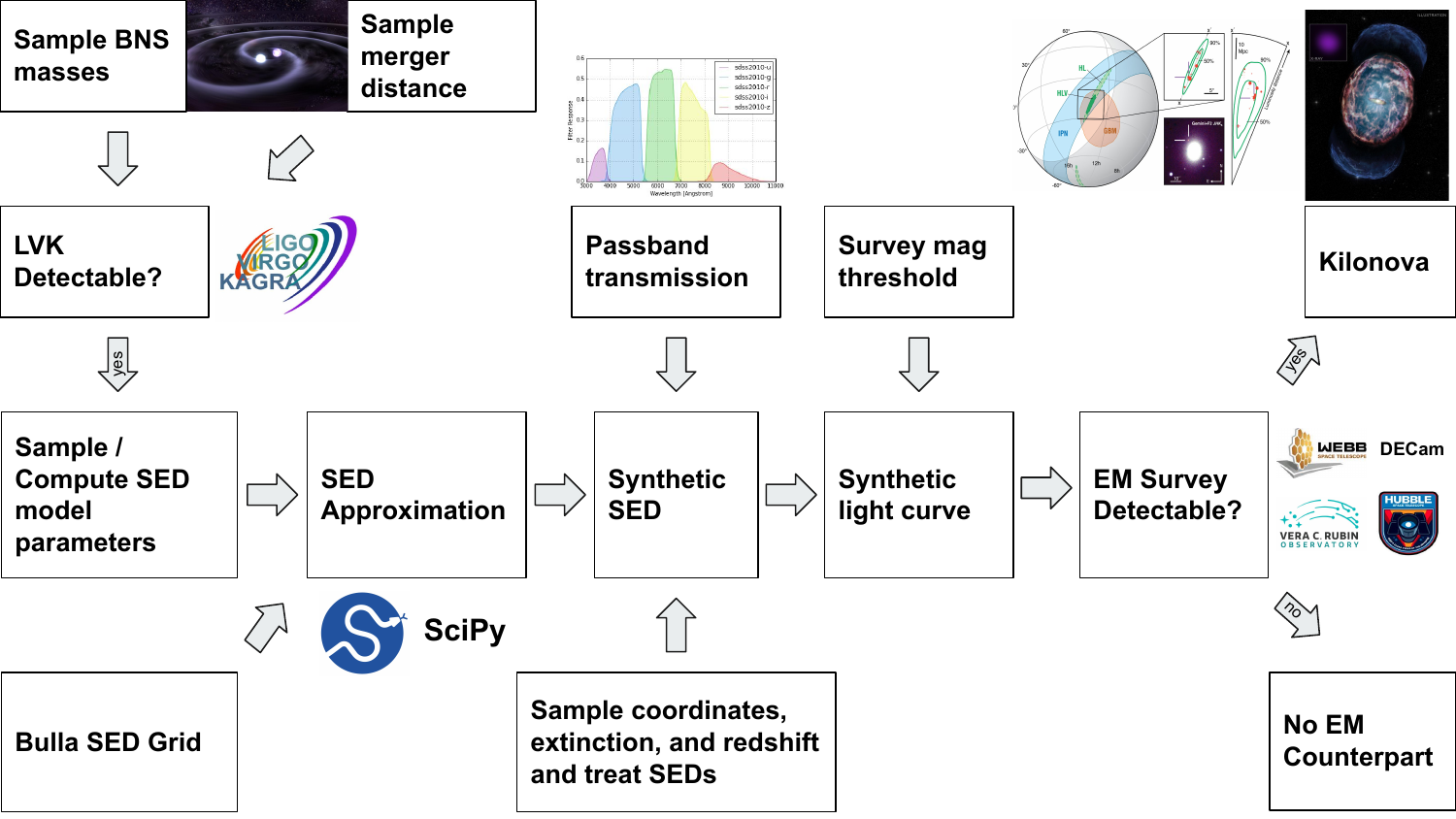}
    \caption{Schematic of the pipeline used to generate synthetic observables for BNS mergers and determine which mergers will produce detectable gravitational waves and electromagnetic counterparts.} 
    \label{fig:pipeline}
\end{figure*}

As the two neutron stars inspiral, they become tidally disrupted, causing neutron-rich material to be ejected from the system. The amount of material ejected depends, among other factors, on the equation of state (EOS)~\citep{Sekiguchi2015,Lattimer2016} being "stiff" or "soft" \citep{Lattimer2016, Shibata2016}.
A neutron star with a stiff EOS exhibits greater pressure, for a given density, and has larger radius causing it to experience greater tidal forces from its companion. In this work we use the SFHo EOS~\citep{Steiner_2013} used in~\citet{Setzer2023} for modeling the kilonova population. Several spectral-energy density (SED) models exist that are parameterized by, for example, the mass and velocity of the ejecta, electron fraction, or opacity \citep[e.g.,][]{Barnes2013,Kasen2015,Metzger_2017,Radice2018}. For this work we use the \textit{bns\_m3\_3comp} model grid developed in~\citet{2019MNRAS.489.5037B, 2020Sci...370.1450D}, henceforth referred to as the ~\citet{2019MNRAS.489.5037B} grid since it has consideration for observing constraints like viewing angles in its parameter space.

SSS2017a or AT2017gfo is the first optically confirmed kilonova from a binary neutron star merger \citep{ 2017Sci...358.1556C,2017ApJ...850L...1L,  2017ApJ...848L..16S, 2017ApJ...848L..24V}, which was detected in conjunction with the gravitational-wave event GW170817~\citep{Abbott_2017:bns_inspiral} and the gamma ray burst GRB170717A~\citep{LIGO_2017:GW_GRB}. This was a landmark discovery for the field of multi-messenger astronomy (MMA) since it was the first detection of a cosmic event via gravitational-waves, a kilonova, and gamma rays.

However, GW170817 remains the only such KN discovery to-date. This is in part due to the current limitations in GW event localization, the coordination required to perform proper follow-up, and the expected rarity of such events. Nonetheless, these events promise many scientific opportunities, such as studying the neutron stars and their EOS \citep{2017Sci...358.1583K, 2017ApJ...850L..19M, 2017ApJ...848L..26S,  2018ApJ...852L..29R, 2019MNRAS.489L..91C,  2020Sci...370.1450D}, understanding r-process nucleosynthesis \citep{2017ApJ...848L..19C, osti_1421833, 2017Sci...358.1570D, 2017ApJ...848L..34M, 2017Sci...358.1574S}, and measuring the expansion of the Universe \citep{Abbott_2018, Abbott_2019, PhysRevResearch.2.022006, 2020Sci...370.1450D}. But, to capitalize on these scientific promises, observers must be prepared to discover and follow-up future BNS events. Understanding the number of observable kilonovae expected during gravitational-wave observing runs would provide critical input for the follow-up efforts within the MMA community.

To address this need, we present here a new methodology to quantify the rates of observable kilonovae during the LVK's ongoing and future observing runs, complimenting similar analysis done recently \citep{2022ApJ...937...79C, 2022ApJ...926..152F,2023arXiv230609234W}. We base our calculation on a number of factors to obtain realistic estimates of these rates (summarized in Figure \ref{fig:pipeline}). First, we sample from the appropriate distributions of BNS masses, astronomical extinction, merger rates, and distances adopted from the literature. We use these sampled parameters, either directly or as inputs to compute flux parameters, to perform interpolation on radiative transfer SED models that we then use to determine the likelihood of electromagnetic counterpart detection. We implement Monte Carlo trials to sample from the distributions in our parameter space to get the distributions of discovery and peak magnitudes, the distances of detected events,  and the number of counterpart detections expected in the LVK O4 and O5 observing runs. The framework is also expandable and can support new parameter models, telescopes, and PSDs from future observing runs can be added as they become available.  

The paper is outlined as follows. In Section \ref{sed}, we detail our usage of existing SED models to build synthetic photometry. In Section \ref{parameters}, we describe the BNS parameter distributions that we use in our analysis. In Section \ref{mc}, we explain our process of sampling from these distributions while accounting for instrumental downtime and other observational constraints. We present the resulting kilonova detection rates in Section \ref{results}.

\section{SED Approximation}
\label{sed}

Running comprehensive, independent simulations \citep{2017Natur.551...80K, 2019MNRAS.489.5037B} to produce SEDs for each merger over all trials is computationally unfeasible. Thus, we use interpolation methods over existing SED grids to approximate the EM radiation. \cite{2019MNRAS.489.5037B} produced a model (\textsc{Possis}) for a grid of kilonovae SEDs simulated using three-dimensional Monte Carlo radiative transfer. These models are parameterised by two different components of the ejecta matter, $m\mathrm{_{ej}^{total}}$: the lanthanide-rich dynamical component, $m\mathrm{_{ej}^{dyn}}$ which is released during the merger and the typically larger, lanthanide-free wind component, $m\mathrm{_{ej}^{wind}}$ released after the merger as a result of unbinding disk matter. Another parameter is the half-opening angle of the lanthanide-rich component of the dynamical ejecta, $\Phi$, and the model has a dependence on the cosine of the observing angle, $\cos \Theta$.

\cite{2020Sci...370.1450D} further improved the model to account for thermal efficiencies and time dependence for the temperature. We choose to use this model in our work. Thus, our SED model is parameterised by $\mathrm{\{ m_{ej}^{dyn}, m_{ej}^{wind}, \Phi, cos \Theta\}}$. It is important to note that  our SED model does not have spin parameter. Although high spin values will have an effect on the resultant kilonova \citep{2021ApJ...922..269R}, the vast majority of Milky Way neutron stars have very low spin \citep{2018PhRvD..98d3002Z}, suggesting that high-spin systems are uncommon. 

\begin{table}
\begin{tabular}{ll}

\toprule
\textbf{Parameter} & \textbf{List of grid values}  \\
\midrule
$\Phi$ &  15, 30, 45, 60, 75   \\
$\cos \Theta$ &  0,  0.1, 0.2, 0.3, 0.4, 0.5, 0.6, 0.7, 0.8, 0.9, 1   \\
$m_{\textrm{ej}}^{\textrm{wind}}$ & 0.01, 0.03, 0.05, 0.07, 0.09, 0.11, 0.13 $M_{\odot}$  \\
$m_{\textrm{ej}}^{\textrm{dyn}}$ & 0.001, 0.005, 0.01, 0.02 $M_{\odot}$ \\
\bottomrule
\end{tabular}
\caption{List of Bulla SED grid values for each of the 4 parameters.}
\label{table:params}
\end{table}

\subsection{Interpolation method}
\label{interpolation-methods}

All SEDs from the above mentioned model, except those with $\Phi = 0$ and $\Phi = 90$, were used to create a 4 dimensional grid. These two $\Phi$ values were excluded as they either lack SEDs for different observing angles or are not available for all permutations of the $m\mathrm{_{ej}^{wind}}$ and $m\mathrm{_{ej}^{dyn}}$.

Table \ref{table:params} describes the discrete points at which SEDs are computed using radiative transfer. While computing the SED for parameters not on the model grid, linear interpolation was used via a regular grid interpolator. The motivation here was that since the flux at every wavelength is known at several finely spaced points in our parameter space through robust simulations, it is reasonable to interpolate between two known points. We use linear interpolation since it is very fast to generate new SEDs on the fly which eliminates the need for pre-computing them, however other interpolation methods with different speed trade-offs also exist within packages like \textsc{Nmma} \citep{pang2022nmma}.

\begin{figure}
    \includegraphics[width=\columnwidth]{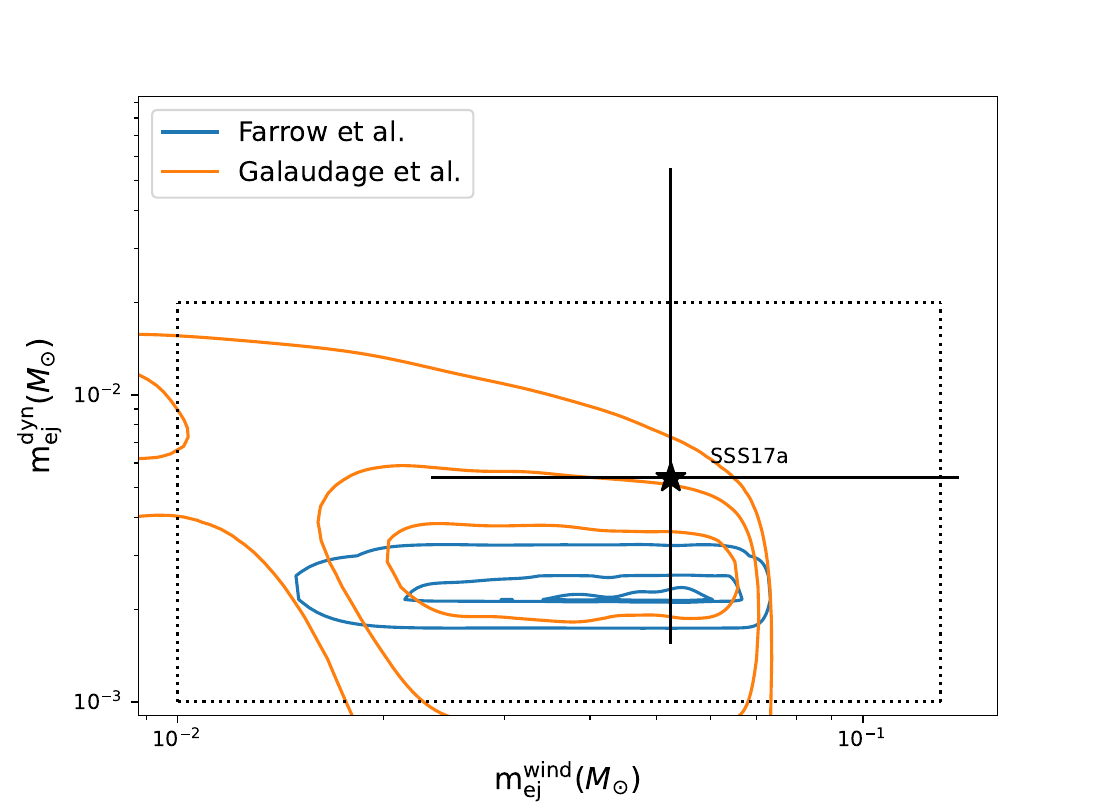}
    \caption{Kernel density estimate contours (corresponding to 20\%, 50\%, and 80\% of the probability mass) for dynamical and wind ejecta mass  from LVK O4 Monte Carlo trials. The dotted lines show the range of grid values for the two components of the ejecta \protect\citep{2019MNRAS.489.5037B}. Data points beyond the grid limits demonstrate the need for an extrapolation method. The SSS17a fit parameters $\left(\mathrm{log_{10}}(m_{\mathrm{ej}}^{\mathrm{wind}}/M_{\odot}) = -1.28_{-0.35}^{+0.42} , \mathrm{log_{10}}(m_{\mathrm{ej}}^{\mathrm{dyn}}/M_{\odot}) = -2.27_{-0.54}^{+1.01}\right)$ were first computed by \protect\cite{2020Sci...370.1450D}. Both the \protect\cite{2019ApJ...876...18F} and \protect\cite{2021ApJ...909L..19G} mass models were used for this analysis.} 
    \label{fig:mc-mej-contour}
\end{figure}

Distributions of the ejecta masses computed during our trials (Figure \ref{fig:mc-mej-contour}) indicate that a non-negligible fraction of binary neutron star mergers will produce $m\mathrm{_{ej}^{dyn}}$ that is greater than the maximum value on the grid ($0.02$ $\mathrm{M_{\odot}}$) or $m\mathrm{_{ej}^{wind}}$ that is less than the minimum value on the grid ($0.01$ $\mathrm{M_{\odot}}$), when sampling component BNS masses from realistic distributions. This necessitates some method for estimating SEDs when the $\mathrm{m_{ej}}$ parameters fall outside the grid range. 

Given the linear relationship between energy radiated and ejecta mass (Section 3.1 \cite{2020FrP.....8..355B}; Equation 4 \cite{1998ApJ...507L..59L}), we have computed scaling laws for the total energy radiated for each $\cos \Theta$ and $\Phi$ pair. If the $m\mathrm{_{ej}^{total}}$ from our BNS merger exceeds the grid limit, we use these linear laws to scale the closest grid SED. 

If our $m\mathrm{_{ej}^{total}}$ is lower than the minimum $m\mathrm{_{ej}^{total}}$ value on the grid, we scale down the closest grid SED using a power law fit since it has the additional benefit of predicting zero flux when the $m\mathrm{_{ej}^{total} = 0}$, according to:
\begin{equation}
    \mathrm{SED = \alpha  SED_{nn}}
\end{equation}
where $\alpha$ is the scaling factor and $\rm SED_{nn}$ is the nearest neighboring SED:
\begin{equation}
    \alpha =
    \begin{cases} 
      \left(\frac{m_{\text{ej}}^{\text{total}}}{ m_{\text{ej-nn}}^{\text{total}}}\right)^n & \text{if } m_{\text{ej}}^{\text{total}} < \text{lowest grid } m_{\text{ej}}^{\text{total}} \\
      
      \left(\frac{m \cdot m_{\text{ej}}^{\text{total}} + c}{m \cdot m_{\text{ej-nn}}^{\text{total}} + c}\right) & \text{otherwise}\\
   \end{cases}
\end{equation}
where $m$ and $c$ denote the slope and intercept for the linear fit respectively and $n$ denotes the exponent for the power law fit. Note that all the best fit scaling parameters (i.e. $m$, $c$, and $n$) were pre-computed for every pair of $(\Phi, \cos \Theta)$. Figure ~\ref{fig:scaling-laws}  shows the best fits for the linear and power scaling laws along with the relative errors for some pairs of $(\Phi, \text{cos } \Theta)$.  Table \ref{table:linear-laws} and table \ref{table:power-laws} document all the parameters for linear and power law scaling respectively.

We use this $m_{\textrm{ej}}^{\textrm{total}}$-dependent interpolation scheme instead of linear extrapolation beyond the regular grid range since a small negative slope over a large extrapolated grid range eventually result in negative fluxes at many wavelengths. These extrapolation artifacts result in non-physical SEDs.

\begin{figure*}
    \includegraphics[width=0.33\linewidth]{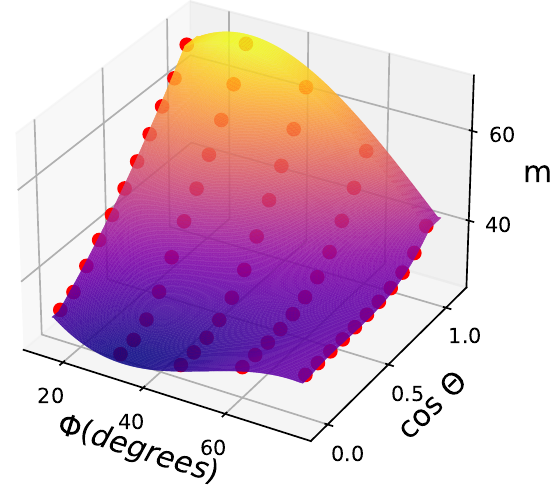}
    \includegraphics[width=0.33\linewidth]{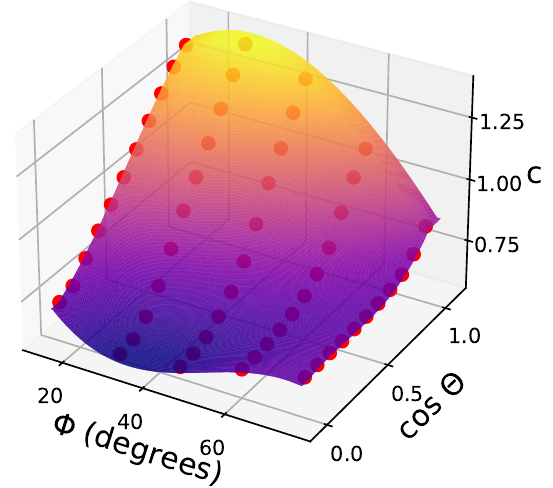}
    \includegraphics[width=0.33\linewidth]{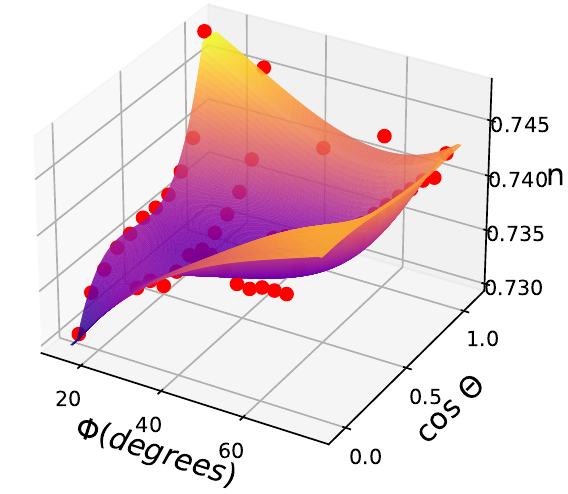}
    \caption{\textbf{Left: } Spline surfaces for the slopes ($m$) of the linear scaling laws for KN SEDs. \textbf{Middle: } Spline surfaces for the intercept ($c$)  of the linear scaling laws. \textbf{Right: } Spline surfaces for the  the exponents ($n$) of the power scaling laws. These laws are used to scale SEDs in cases where the ejecta masses exceed the grid limits of the SED model. } 
    \label{fig:spline-fits}
\end{figure*}

Since we want to sample the $(\Phi, \cos \Theta)$ parameters from a continuous range rather than the discrete points computed above to avoid quantization, we fit a spline function to our data for all three scaling parameters (namely $m$, $c$, and $n$) using the smooth bivariate spline. Figure \ref{fig:spline-fits} shows the spline surfaces fit to the discrete points. We compute our scaling parameters from this surface for all our Monte Carlo trials. The sum of residuals from the surfaces are provided in Table \ref{table:residuals}.

Using this piecewise extrapolation method ensures that we can always get SEDs that have reliable total flux since we are scaling the SEDs based on the ejecta mass. However, this method fails to take into account any changes in color as a function of the $m\mathrm{_{ej}^{total}}$ since there was no obvious statistical trend for how the spectrum shifted. Doing this correctly will require updating the original radiative transfer simulations for a larger range of $m\mathrm{_{ej}}$ values which is outside the scope of this paper.

\subsection{Redshift and Extinction}

Since most binary neutron star mergers are expected to be of extragalactic origin, we treat our SEDs for both host and Milky Way extinction. For the host galaxy, considered to be at rest, we use the CCM89Dust effect based on work from \cite{1989ApJ...345..245C}. The $E_{B-V}^{\mathrm{host}}$ is computed for each SED using the $A_V$ sampled from the distribution described in Equation \ref{eq:av} and $R_V = 3.1$ using the following equation
\begin{equation}
    E_{B-V}^{\mathrm{host}} = \frac{A_V}{R_V}
\end{equation}

For the Milky Way galaxy, considered to be the observing frame, we use the F99Dust effect to redden the SED based on work from \cite{1999PASP..111...63F}. Here, we use the SFD dust map \cite{1998ApJ...500..525S} to find the $E_{B-V}^\mathrm{MW}$ based on the RA and Dec of each event.

Finally, we redshift the SED based on the luminosity distance of our kilonova. Details about how this distance is sampled are provided in the parameter distribution section (Section \ref{parameters}). All effects are applied to the SED within \textsc{Sncosmo} \citep{sncosmo}.

\section{Parameter distribution}
\label{parameters}
Our Monte Carlo simulations use the aforementioned pipeline to estimate SEDs and produce the associated light curves. We sample the component neutron star masses in order to determine if the merger can be detected via gravitational waves. The component masses are also used to compute the $m\mathrm{_{ej}^{wind}}$ and $m\mathrm{_{ej}^{dyn}}$ which, in conjunction with the sampled $\Phi$ and $\cos \Theta$, are used to estimate the SEDs. Finally we sample $A_V$ and the event coordinates to treat the SEDs and generate the synthetic observables. The distributions for all of these inputs are described below. Table \ref{table:input-parameters} summarizes these distributions.

\subsection{BNS mass distribution and computing ejecta mass}

For our BNS pairs, we consider the standard formation scenario where binaries consist of a first-born recycled neutron
star sped up from accretion (with mass $M\mathrm{_{recycled}}$) and a second-born slow neutron star (with mass $M\mathrm{_{slow}}$).

Analyzing the the mass distributions of these two distinct NS populations has been the subject of  numerous studies. \cite{2019ApJ...876...18F} used a two-peak Gaussian for the recycled NS (Table \ref{table:exg-gaussian}) and a flat distribution with the range $[1.16, 1.42]$ $\mathrm{M_{\odot}}$ for the slow NS, with further analysis done by \cite{ 2022ApJ...926...79G}. 
 \cite{2021ApJ...909L..19G} 
used a two-peak Gaussian to describe both the slow and recycled NS mass distributions (Table \ref{table:exg-gaussian}). The motivation behind the two peak Gaussian model for slow neutron stars is to reconcile the disagreement between the empirical galactic data on BNS pairs from radio sources \citep{2019ApJ...876...18F} and the distribution that would be required to explain gravitational wave events like GW190425 \citep{2020ApJ...892L...3A}. It is worth noting that both results use the same two peak Gaussian to explain the recycled NS distribution. 

\begin{figure}
    \includegraphics[width=\columnwidth]{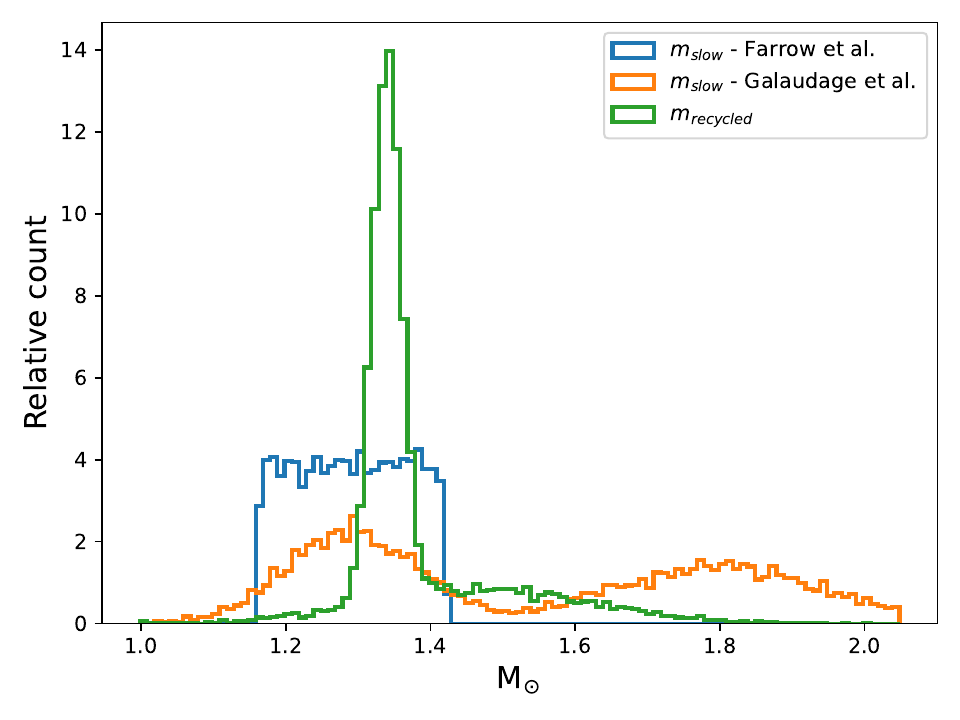}
    \caption{Distribution of sampled masses of BNS populations for LVK O4 and O5 simulation. The uniform distribution for the slow NS (blue) is from  \protect\cite{2019ApJ...876...18F}. The two peak Gaussian for the slow NS (orange) is from \protect\cite{2021ApJ...909L..19G}. The distribution for the recycled NS (green) is shared by both models. } 
    \label{fig:bns_mass}
\end{figure}

Our simulations support three different BNS mass models: the \cite{2019ApJ...876...18F} and \cite{2021ApJ...909L..19G} models discussed above (Figure \ref{fig:bns_mass}) and a flat distribution with some astrophysical priors from the \textsc{Kilopop} package \citep{Setzer2023}. While presenting the results, we only use the \cite{2019ApJ...876...18F} and \cite{2021ApJ...909L..19G} models since the uniform distribution is not empirically motivated. 
Ultimately, we find that while the choice of mass distribution changes the distribution of ejecta properties, it makes little difference in the number of dicoverable KNe estimated by our simulations (Table \ref{table:final-summary}).

\subsubsection{Discussion of EOS}

In addition to the components masses, the equation of state employed will also have an impact on the both the $M\mathrm{_{TOV} }$ 
 \citep{PhysRev.55.374} and the ejected matter. Consequently, the post merger remnant and the resulting kilonova also depend on the assumed EOS. A neutron star with a stiff EOS exhibits greater pressure,
for a given density, and has larger radius causing it to experience
greater tidal forces from its companion. This results in greater dynamical ejecta. The post-merger remnant influences the opacity of the ejecta \citep{Kasen2015, Radice2018}. A long-lived neutron star remnant will emit neutrinos, causing the electron fraction to increase and thus reduce the production of lanthanides. In the case of GW170817, \cite{Margalit_2017} were able to constrain the nature of the merger remnant to be a short-lived hyper-massive neutron star through the gravitational wave and electromagnetic signals.  
 
While there is still no clear EOS that is most favorable,  significant work concerned with fitting functions \citep{Setzer2023} to ejecta properties is done using the SFHo EOS \citep{2013ApJ...774...17S}. Since we make use of these fitting functions in our work, we opt to use the same EOS. It is important to note that the equation of state  used in this work places constraints on both minimum and maximum masses for neutron stars. For this reason, we cut off the tails of our mass distributions at $1 \mathrm{M_{\odot}}$ and $2.05 \mathrm{M_{\odot}}$ respectively. 

There is some evidence that the SFHo EOS results in low $M\mathrm{_{TOV}}$ compared to empirical data \citep{2020MNRAS.494..190F} and thus may not capture the  population diversity. However, recomputing the ejecta fits for different EOS models is outside the scope of this paper.

\begin{table}
\begin{tabular}{lll}

\toprule
\textbf{Parameter} & \textbf{Recycled} & \textbf{Slow}  \\
\midrule
$\zeta$ & 0.68 & 0.5   \\
$\mu_{1}$ &  $1.34 M_{\odot}$ & $1.29 M_{\odot}$  \\
$\sigma_{1}$ &  $0.02 M_{\odot}$ & $0.09 M_{\odot}$  \\
$\mu_{2}$ & $1.47 M_{\odot}$  & $1.8 M_{\odot}$  \\
$\sigma_{2}$ & $0.15 M_{\odot}$  & $0.15 M_{\odot}$ \\
$M_{\textrm{low}}$ & ---& $1.16 M_{\odot}$ \\
$M_{\textrm{high}}$ & ---& $1.42 M_{\odot}$ \\
\bottomrule
\end{tabular}
\caption{BNS population parameters. $\zeta$ defines the fraction of binaries in the low mass peak.}
\label{table:exg-gaussian}
\end{table}

\subsubsection{Computing ejecta masses}

While the two NS masses are not themselves parameters for the chosen SED model, we use the masses to compute the ejecta parameters. Although we are not introducing any new ejecta fits in this work, we use this section to recapitulate the methodology for computing these ejecta parameters.

\cite{Setzer2023} used the fitting function introduced by \cite{2019MNRAS.489L..91C} and data from 259 Numerical Relativity (NR) simulations \citep{2018ApJ...869..130R} to 
come up with the final fitting function  for the dynamical ejecta mass which depends on the masses ($\mathrm{M_{1,2}}$) and compactness ($\mathrm{C_{1,2}}$) of the merging neutron stars and is given by
\begin{equation}
    \label{eq:mej_dyn}
    \mathrm{log_{\textrm{10 }}} (m\mathrm{_{ej}^{\textrm{dyn fit}}}) = {\left[ a  \frac{(1 - 2 C_1)M_1}{C_1} + b  M_2  \left(\frac{M_1}{M_2}\right)^n + \frac{d}{2} \right] + [1\leftrightarrow2]},
\end{equation}
where $a = -0.0719$, $b = 0.2116$, $d = -2.42$, and $n = -2.905$, and $[1\leftrightarrow2]$ refers to
repetition of the preceding fit with the
indices interchanged. The compactness of the neutron star is given by is given by
\begin{equation}
    \label{eq:compactness}
    C = \frac{G M}{c^2 R}.
\end{equation}
Due to the logarithmic nature of our fit, two heavy neutron stars ($\mathrm{\sim 2 M_{\odot}}$ each) result in dynamical ejecta mass that can exceed $\mathrm{1 M_{\odot}}$. \cite{Metzger-KN} (Section 3.1.1) and work referenced within find that the total dynamical ejecta from BNS mergers lie in the range $\mathrm{10^{-4} M_{\odot} - 10^{-2} M_{\odot}}$. Thus, we limit the maximum dynamical ejecta from our mergers to $\mathrm{0.09M_{\odot}}$ which means our final $m\mathrm{_{ej}^{dyn}}$ is
\begin{equation}
    \label{eq:mej_wind}
    m\mathrm{_{ej}^{dyn}} = \textrm{min} \left( 0.09, m\mathrm{_{ej}^{\textrm{dyn fit}}} \right).
\end{equation}
The other ejecta parameter, $m\mathrm{_{ej}^{wind}}$, is some fraction of the disk mass, $m\mathrm{_{disk}}$, where the two are related by
\begin{equation}
    \label{eq:mej_wind}
    m\mathrm{_{ej}^{wind}} = \zeta  m\mathrm{_{disk}},
\end{equation}
and $\zeta$ is the unbinding efficiency which is sampled uniformly from the range 10--40\% and $m\mathrm{_{disk}}$ is computed as follows
\begin{equation}
    \label{eq:m_disk}
    \mathrm{log_{10}} (m\mathrm{_{disk}) = max}{ \left(-3, a \left(1 + b \tanh \left[ \frac{c - (M_1 + M_2)/M\mathrm{_{thr}}}{d} \right] \right) \right)},
\end{equation}
where $a = -31.335$, $b = -0.9760$, $c = 1.0474$, and $d = 0.05957$.

Finally, $M\mathrm{_{thr}}$, the mass
threshold for prompt black hole collapse, is computed using the following \citep{2013PhRvL.111m1101B},
\begin{equation}
    \label{eq:m_disk}
    M\mathrm{_{thr}} = \left( 2.38 - 3.606 \frac{M\mathrm{_{TOV}}}    
    {R_{1.6\mathrm{M_{\odot}}}} \right) M\mathrm{_{TOV}}.
\end{equation}
If ${m_1 + m_2 \geq M\mathrm{_{thr}}}$, then we set both dynamical and wind ejecta to zero, ensuring there is no luminous remnant.

\subsection{$A_{V}$ distribution}

Modeling the extinction of kilonovae host galaxies based on empirical data is unlikely to yield accurate results given the single data point - host NGC 4993 for GW170817 \citep{2017ApJ...848L..30P}. For this reason, we sample from an extinction distribution for galaxies known to host supernovae. 

\cite{2009ApJS..185...32K} (Equation 18) inferred the mean reddening parameter for host galaxies using a SDSS-II sample of Type 1a supernovae and found that their extinction can be well-explained by an exponential function of the form
\begin{equation}
    \label{eq:av}
    P (A_V) = \exp \left(\frac{-A_V}{ \tau_V}\right),
\end{equation}
where $\tau_V = 0.334 \pm 0.088$. We sample from this distribution with a fixed $\tau_V = 0.334$. 

\subsection{Spatial and distance distributions}

For each set of trials, we first define a cube of length $l$ within which we simulate the events. We sample $x$, $y$, and $z$ coordinates from a uniform, random distribution in the range of [$-\frac{\mathrm{l}}{2}$,  $+\frac{\mathrm{l}}{2}$ ] where $l$ is specific to the simulation and detailed in Section \ref{mc}. The Cartesian coordinates are converted to spherical coordinates to compute the event $\text{RA}$ and $\text{Dec}$ values. The euclidean distance is computed using the following:
\begin{equation}
    D = \textrm{0.05 Mpc} + \sqrt{x^2 + y^2 + z^2}
\end{equation}
The additional $0.05\, Mpc$ term is added to ensure a minimum distance for events. We use these distances to compute the redshifts, assuming a flat Lambda-CDM cosmology \footnote{https://docs.astropy.org/en/stable/api/astropy.cosmology.FlatLambdaCDM.html}. Since we are only simulating mergers $\le500$~Mpc, sampling distances is roughly equivalent to sampling redshifts.

\subsection{$\Phi$ distribution}

$\Phi$ describes the half-opening angle at which the lanthanide-rich component of the dynamical ejecta is distributed  close to the equatorial plane during the merger giving rise to the  red kilonova. The remainder of the ejecta is expelled further from the equator resulting in the blue kilonova. The lack of comprehensive data on kilonovae means that we do not have reliable distributions for $\Phi$. For this reason, we choose to  sample from a uniform continuous distribution of values for $\Phi$ in the range $[15, 75]$, the entire range for values for which complete simulations exist (See Section \ref{interpolation-methods}).

\subsection{$\cos \Theta$ distribution}

Given the random distribution of BNS mergers in space, it is safe to assume that the  cosine of the observing angle will be uniformly-distributed. Specifically, we sample from a continuous distribution of $\cos \Theta$ values in the range $[0, 1]$.

We also use the observing angle to compute the inclination, defined as the angle between the line of sight and the total angular momentum, \citep{2019PhRvX...9c1028C} ($\Omega$) which is used as a parameter in the GW waveform generation (Section \ref{sec:gw-det}).

\begin{equation}
    \Omega = \textrm{min} (\Theta, 180 - \Theta)
\end{equation}

\begin{figure}

    \includegraphics[width=\columnwidth]{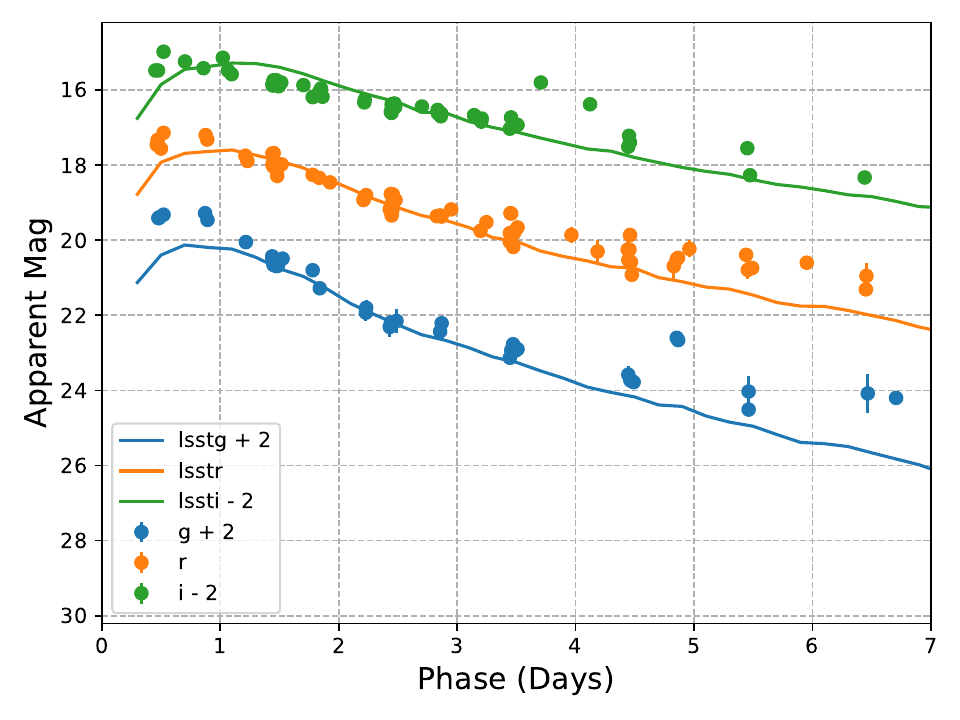}
    \caption{Light curves for GW170817 constructed using interpolated spectra with fit parameters computed by \protect\cite{2020Sci...370.1450D}, overplotted with real photometry. Appendix \ref{photometry-credit} provides a complete list of the sources for all the photometry used in this plot.} 
    \label{fig:GW170817LC}
\end{figure}

\begin{table}
\begin{tabular}{ll}

\toprule
\textbf{Parameter} & \textbf{Distribution}  \\
\midrule
$A_V$ &  $\exp \left(\frac{-A_V}{ \tau_V}\right)$ \\
$\Phi$ &  $\mathrm{U}(15, 75)$ \\
$\cos \Theta$ &  $\mathrm{U}(0, 1)$ \\
\bottomrule
\end{tabular}
\caption{Table of input parameters distributions used for LVK O4.}
\label{table:input-parameters}
\end{table}

\section{Monte Carlo Trials}
\label{mc}

For each trial, we sample the BNS component masses, distances, coordinates, $A_V$, $\Phi$, and $\cos  \Theta$ from the distributions mentioned above. The criteria used for determining gravitational wave and electromagnetic detections are specified below. We found that the expected numbers of discoverable kilonovae begin to converge after a few hundred iterations of our simulations for both the O4 and O5 observing runs. For this reason, we run all our simulations for 1000 iterations since we do not expect the results to change significantly with further increase in the number of trials. 

\subsection{Finding the number of events}

For each trial, we first need to find the number of events, henceforth called $n\mathrm{_{events}}$ where:

\begin{equation}
    n_\mathrm{events} = \mathrm{rate \cdot volume \cdot time}
\end{equation}

The time is computed from the overlap duration of our chosen optical survey and the LVK observing run. Based on Table \ref{table:hd}, we know that the maximum distances for BNS GW detections is $\sim 253$ Mpc and $\sim 449$ Mpc for LVK observing runs O4 and O5, respectively. Thus, the lengths of our event cube for simulations, $l$, are set to $510$ Mpc and $910$ Mpc respectively. This ensures that the limiting factor for kilonova discovery is always either the sensitivity of the LVK detectors or the limiting magnitude of our survey.

\begin{figure}
    \includegraphics[width=\columnwidth]{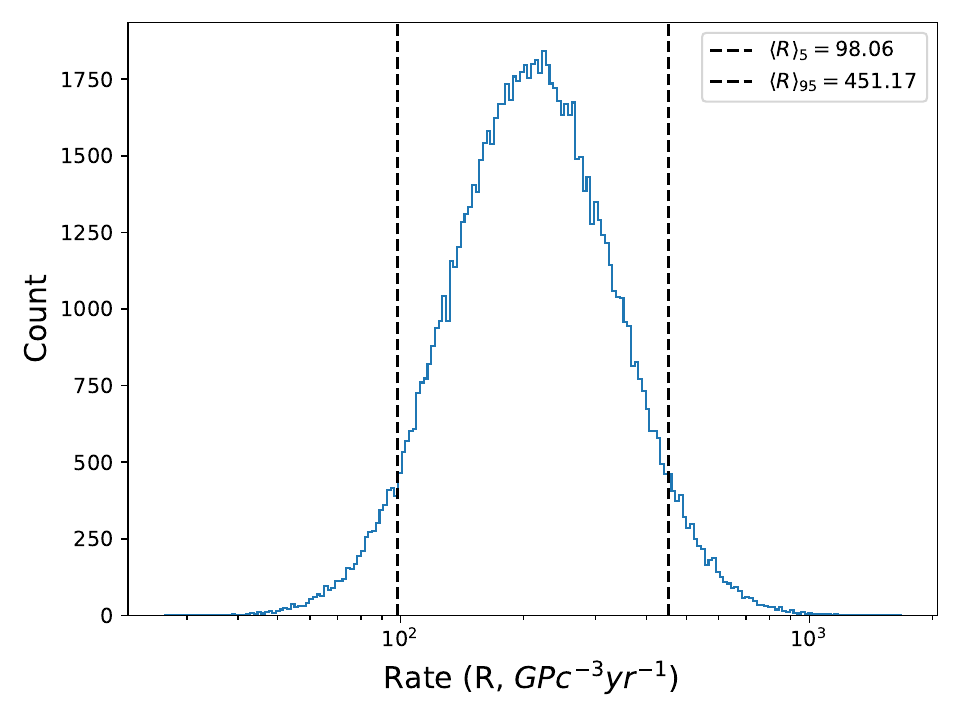}
    \caption{Distribution of BNS merger rates approximates the log normal distribution mentioned in the LVK user guide.} 
    \label{fig:merger_rates}
\end{figure}

The merger rate model is another configurable option in our simulations. Considerable work has been done to understand the frequency of BNS mergers \citep{2021arXiv211206878N, 2023PhRvX..13a1048A}. For this work, we set the BNS merger rate to $\mathrm{210^{+240}_{-120}  Gpc^{-3} yr^{-1}}$ with log-normal uncertainties, in accordance with the LVK user guide \footnote{https://emfollow.docs.ligo.org/userguide/capabilities.html \label{url:user-guide}}(Figure \ref{fig:merger_rates}) \citep{2023PhRvX..13a1048A}.

\subsection{Detecting gravitational waves from mergers}
\label{sec:gw-det}

\begin{figure}
    \includegraphics[width=\columnwidth]{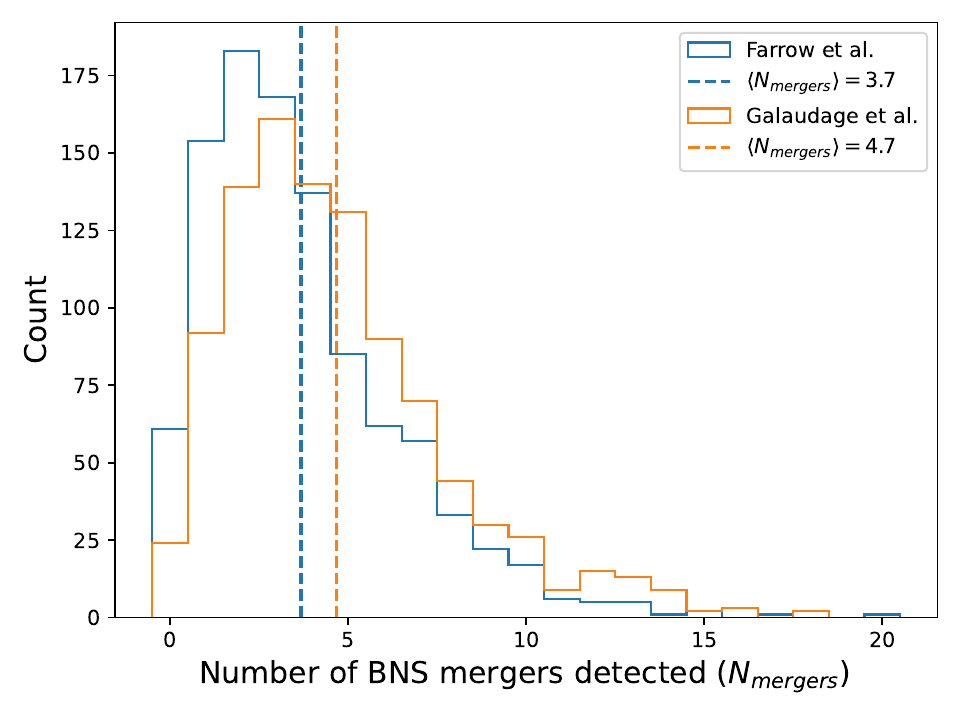}
    \includegraphics[width=\columnwidth]{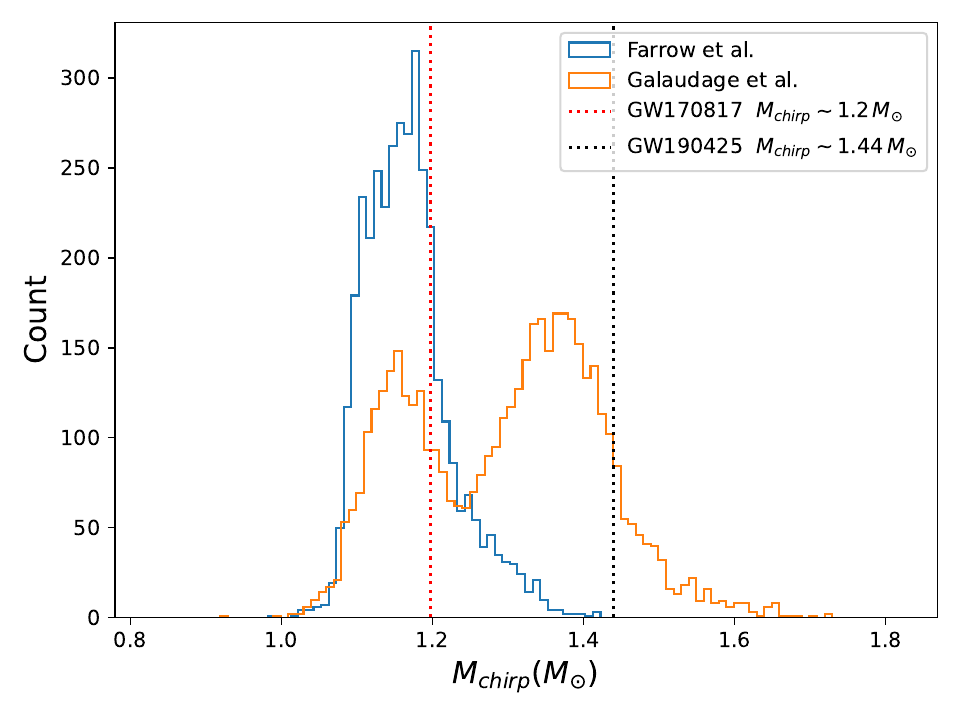}
    \includegraphics[width=\columnwidth]{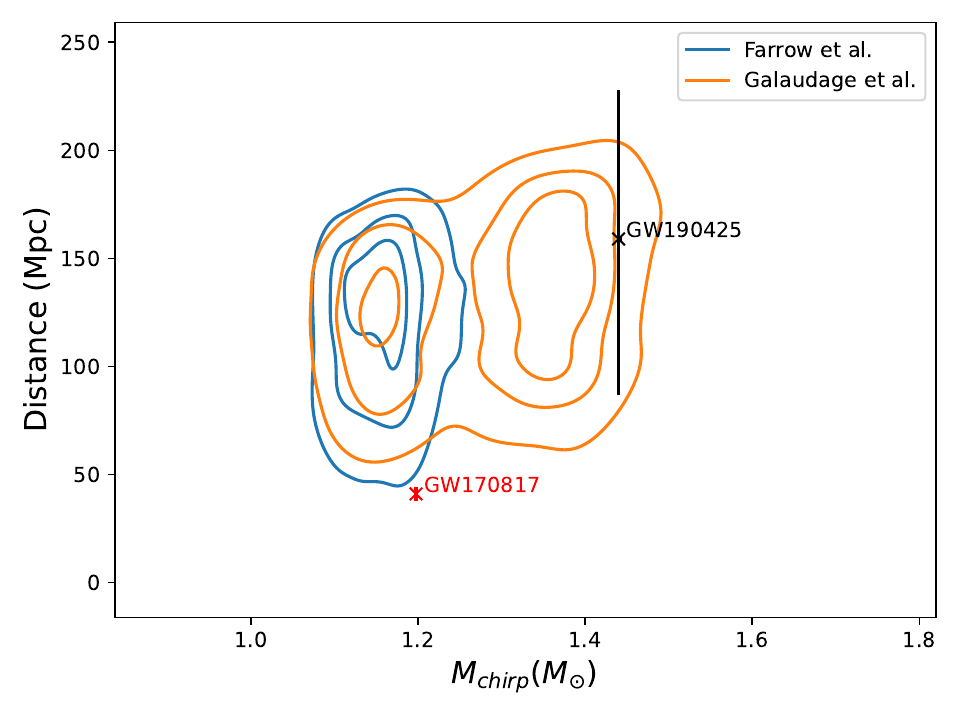}
    \caption{Predictions for BNS mergers detected by LVK during O4 using either the \protect\cite{2019ApJ...876...18F} (blue curves) or \protect\cite{2021ApJ...909L..19G} (orange curves) mass models.  \textbf{Top: } Distribution of the number of GW-detected BNS mergers for our Monte-Carlo trials.  The average number of mergers is represented by a vertical dashed line. \textbf{Middle: } Distribution of chirp masses for all BNS mergers with GW detections. \textbf{Bottom: } Kernel density estimate contours (corresponding to 20\%, 50\%, and 80\% of the probability mass) for the luminosity distance as a function of chirp mass for all BNS mergers with GW detections.   GW170817 and GW190425 are represented by red and black points (dashed vertical lines), respectively, in the bottom (middle) panel.}
    \label{fig:O4-bns_mergers} 
\end{figure}

The maximum distance values at which a BNS merger is detectable is a function of the component masses of our binary system and its inclination~\citep{2021CQGra..38e5010C}. 

Once we sample our mass distributions to find the component masses, we compute its gravitational waveform which acts as our signal. This waveform is parameterized by the masses of our coalescing binary and the inclination angle, $\Omega$. Since we are dealing with binary neutron stars, we use the TaylorF2 waveform \citep{2009PhRvD..80h4043B, 2019PhRvD..99l4051M}, which assumes neutron stars are non-spinning point masses and has been used for BNS merger rate modeling before \citep{2021arXiv211206878N}. Given that we only use the waveform to determine the maximum distance at which a detector would be able to discover a merger, the TaylorF2 waveform is sufficiently accurate (compared to a more comprehensive model like IMRPhenomPv2\_NRTidal) and much faster to compute, a key advantage for the speed of our Monte Carlo trials. 

For each instrument, we use the PSD$^{\ref{url:user-guide}}$ which describes the noise at a given frequency. We then use a signal to noise ratio threshold of 8 (to remain consistent with LVK$^{\ref{url:user-guide}}$) \citep{2022ApJ...924...54P} to determine the maximum distance at which such a merger would produce a detection. Since we already know the luminosity distances for each of our mergers, we can determine if the event would produce a detection for each of the instruments.

Another aspect to take into account is the correlation in uptimes between the LVK detectors and their respective duty cycles. We used data on the correlation between different detectors from the LIGO O3a run\footnote{https://gwosc.org/detector\_status/O3a/} to create a correlation matrix where the rows and columns are ordered by LIGO Hanford, LIGO Livingston, Virgo, and KAGRA respectively. Since we do not have duty cycle correlation data for KAGRA during O3, we assume the same $\sim 56 - 58 \%$ correlation as Virgo. These values are consistent with the current $\sim 58 \%$  duty cycle correlation between LIGO - Livingston and LIGO - Hanford  reported for LVK O4 during the September 21, 2023 LVEM call \footnote{https://wiki.gw-astronomy.org/OpenLVEM/Telecon20230921}.

\begin{equation}
\mathrm{COR}=
  \begin{bmatrix}
    1.0 & 0.56 & 0.56 & 0.56 \\
    0.56 & 1.0 & 0.58 & 0.58 \\
    0.56 & 0.58 & 1.0 & 0.56 \\
    0.56 & 0.58 & 0.56 & 1.0 \\
  \end{bmatrix}
\end{equation}

We use this to create a detector uptime correlation matrix of dimension $4 \times 4$. For each trial we also create a matrix of random numbers in the range [0,1] of dimension $n\mathrm{_{events} \times 4}$. We multiply the random numbers with the correlation matrix and scale all values to the range [0,1], using a min-max scaler, resulting in a matrix of dimensions $n\mathrm{_{events} \times 4}$. Each column of this matrix is a series that represents the probability the detector is active when each merger in the trial is taking place. Since we already know the duty cycles for each detector (Table \ref{table:duty-cycles}), we can set the detector to observe during an event if this probability during the event is less than the value of the duty cycle for that instrument. We can repeat this for all four detectors and determine which instruments would be on during each of our mergers in a given trial. 

If a detector is on and observing during a merger and the merger is within the detection range for the component masses at their inclination, then we consider the event to be detected since we are not modeling any coincidental terrestrial noise, antenna patterns on the detector, or accounting for software failures. We then determine how many instruments will detect the merger by checking each detector's status and its detection capability.

It is important to note that non-detections, in the context of GW events, can encode vital information that can help improve sky map localization. For instance, the non-detection of GW170817 by Virgo, despite observing during the event, helped narrow down the localization \citep{LIGOScientific:2017vwq}. However, since we are not generating sky maps for this work, we choose to ignore antenna patterns while considering detections.  

A duty cycle of 70\% was used based on the observing capabilities as reported in the LVK Userguide$^{\ref{url:user-guide}}$. However, both Virgo and KAGRA will not be operating for the entirety of the 18 month period, so we encode this information into their duty cycles.

\begin{equation}
    \mathrm{\text{duty cycle} = 0.7  \frac{\text{operating months}}{18}}
\end{equation}

Based on the latest observing plan\footnote{https://observing.docs.ligo.org/plan/} available at this time (dated October 14, 2023), we assume that Virgo will operating at optimum sensitivity for $12$ months and KAGRA for $7$ months. Table \ref{table:duty-cycles} details the duty cycles for all 4 detectors in the LVK network used in our simulations.

At this stage of the pipeline, we already have data on the BNS mergers that were detected in our simulation. Fig \ref{fig:O4-bns_mergers} shows the properties of these mergers. We find that the median number of GW detections for merging neutron stars over LVK O4 is $\sim 3-4$, depending on the mass model used.

\begin{table}
\begin{tabular}{lllll}

\toprule
\textbf{Detector} & \textbf{Operating Months} & \textbf{O4 Duty cycle}   \\
\midrule
LIGO - Livingston &  18 &  0.7  \\
LIGO - Hanford  &  18 &  0.7  \\
Virgo  &  12 &  0.47  \\
KAGRA  &  7 &  0.27  \\
\bottomrule
\end{tabular}
\caption{Duty cycles for detectors used for the LVK O4 Monte Carlo simulations}
\label{table:duty-cycles}
\end{table}

\subsection{Detecting EM counterparts from mergers}

As mentioned in the Section \ref{parameters}, we can find the $m\mathrm{_{ej}^{wind}}$ and $m\mathrm{_{ej}^{dyn}}$ for each merger. If the $m\mathrm{_{ej}^{total} > 0}$, then we conclude that the merger has left behind a kilonova. Thus, we use the SED approximation method described in Section \ref{sed} to produce synthetic light curves for all mergers that have non-zero $m\mathrm{_{ej}^{total}}$. We can use the synthetic photometry and the survey's detection thresholds to find the discovery magnitude and peak magnitude of each detectable kilonova.

Next, we need to account for the fraction of events that would be lost to light from the sun, called $F\mathrm{_{sun\, loss}}$. This value changes for different surveys and is a configurable option. We uniformly sample a number between 0 and 1 for each of the $n\mathrm{_{events}}$; if this number is greater than $F\mathrm{_{sun\, loss}}$, then the event is not lost to the sun.

Finally, we label an event as $n\mathrm{_{good}}$ if it was detected by $n$ GW detector(s), has non-zero ejecta, has a peak magnitude that can be detected by the survey, and is not lost to the sun since these filters provide all the necessary, but not sufficient, conditions for the discovery of the kilonova. Based on our simulations, typically $\sim 1-3 \%$ of the BNS merger have a prompt collapse to black holes, resulting in zero ejecta.

Since there are 4 detectors in the LVK network  we will be doing our analysis for n = 1, 2, 3, and 4. It is worth noting that a single instrument detection will typically yield very poorly localized sky maps (on the order of ten thousand sq degrees) and lower network SNR which makes targeted search for kilonovae difficult. 

This concludes the entire methodology we use in order to estimate the rate of discoverable BNS kilonovae.

\begin{figure}
    \includegraphics[width=\columnwidth]{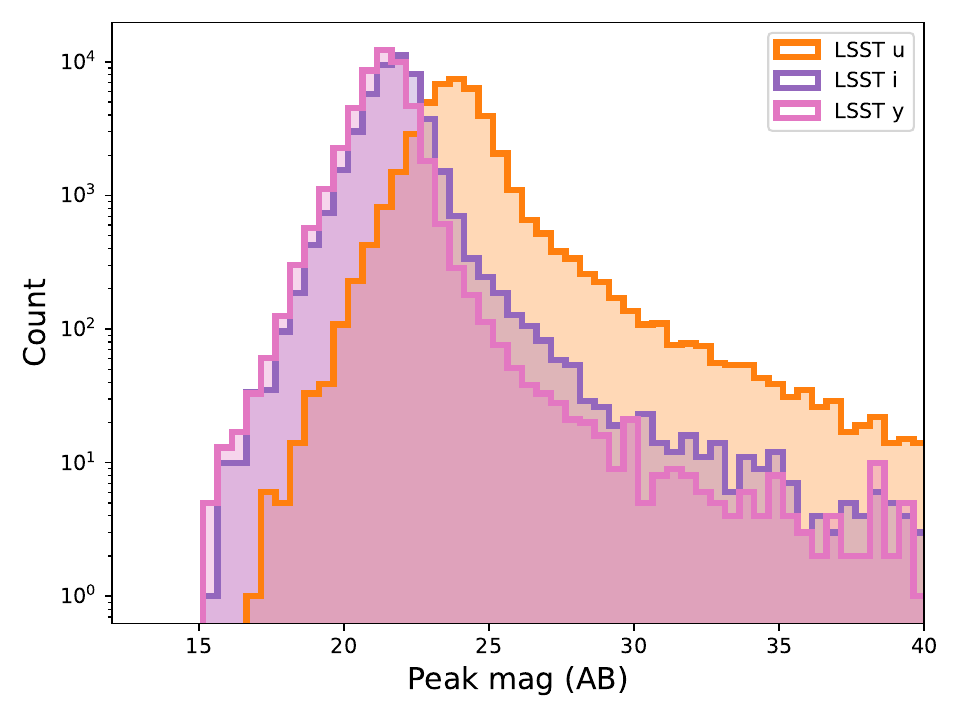}
    \caption{Distribution of apparent peak magnitude for all kilonovae simulated, including those without an EM or GW detection (up to 40 magnitudes) for LVK O4.} 
    \label{fig:kn_mags}
\end{figure}

\section{Results}
\label{results}

In this section we will discuss the results of our Monte Carlo trials. It is important to note that these results indicate the best case scenario for kilonova detection since they do not account for observing inefficiencies (like weather, tiling etc.) or poor gravitational wave skymap localizations.

\begin{table*}
\begin{tabular}{ccccccccccc}

\toprule
\textbf{Run} &
\textbf{Survey} &
\textbf{BNS mass model}  &
\textbf{GW only}  &
\textbf{1 GW + EM}  &
\textbf{2 GW + EM}  &
\textbf{3 GW + EM}  &
\textbf{4 GW + EM}  &
\textbf{All GW + EM}  &
$\mathop{\mathbb{E}}$ \textbf{(KN)}\\
\midrule

O4 &
DECam&
\cite{2021ApJ...909L..19G} &
${ 4}_{- 3}^{+ 7}$&
${ 0 }_{- 0 }^{+ 2 }$ &
${ 1 }_{- 1 }^{+ 3 }$ &
${ 0 }_{- 0 }^{+ 1 }$ &
--- &
${ 2 }_{- 2 }^{+ 3 }$ &
2.2 &
\\
\\

 &
 &
\cite{2019ApJ...876...18F} &
${ 3 }_{- 3}^{+ 6}$ &
${ 0}_{- 0}^{+ 2}$ &
${ 1 }_{- 1}^{+ 2}$ &
${ 0 }_{- 0}^{+ 1}$ &
--- &
${ 1}_{- 1}^{+ 4}$ &
1.8 &
\\
\\

O5 &
LSST&
\cite{2021ApJ...909L..19G}  &
${ 42 }_{- 25}^{+ 47}$ &
${ 5 }_{- 4 }^{+ 7 }$ &
${ 13 }_{- 8 }^{+ 16 }$ &
${ 0 }_{- 0 }^{+ 2 }$  &
${ 1 }_{- 1 }^{+ 2 }$ &
${ 19 }_{- 11}^{+ 24 }$ &
21.6 &\\ \\

 &
 &
\cite{2019ApJ...876...18F}  &
 ${ 42 }_{- 24}^{+ 54}$ &
 ${ 5 }_{- 4 }^{+ 8 }$ &
 ${ 13 }_{- 9 }^{+ 16 }$ &
 ${ 0 }_{- 0 }^{+ 2}$ &
 ${ 1 }_{- 1 }^{+ 2 }$ &
 ${ 19 }_{- 11 }^{+ 24}$ &
 21.7 &\\

\bottomrule
\end{tabular}
\caption{Summary of kilonova discovery estimates over LVK O4 and O5 from this work. Last column presents the expected number for KNe. All other columns in this table represent the middle 90\% credible intervals with the median, $5^{th}$, and $95^{th}$ percentile numbers being reported.}
\label{table:final-summary}
\end{table*}

\begin{table}
\textbf{Mean luminosity distances (in Mpc)}\\ \\
\begin{tabular}{ccccccc}

\toprule
\textbf{Run} &
\textbf{BNS mass model}  &
\textbf{1 GW}  &
\textbf{2 GW}  &
\textbf{3 GW}  &
\textbf{4 GW}  &\\
\midrule

O4 &
\cite{2021ApJ...909L..19G} &
124 &
130 &
83 &
--- &
\\
\\

 &
\cite{2019ApJ...876...18F} &
118 &
122 &
76 &
--- &
\\
\\

O5 &
\cite{2021ApJ...909L..19G} &
228 &
231 &
110 &
89 &
\\
\\

 &
\cite{2019ApJ...876...18F} &
228 &
231 &
108 &
90 &\\
\bottomrule
\end{tabular}
\caption{Summary of mean luminosity distances (in Mpc) of discoverable KNe for LVK O4 and O5 based on simulations with 1, 2, 3, or 4 coincidental GW detection(s) and an EM detection. The missing statistics for the 4-detector events during LVK O4 is due to a negligible number of mergers having coincidental detections on 4 instruments.}
\label{table:final-dist}
\end{table}

\begin{table}
\textbf{Mean peak magnitudes (AB)}\\ \\
\begin{tabular}{ccccccc}
\toprule
\textbf{Run} &
\textbf{BNS mass model}  &
\textbf{1 GW }  &
\textbf{2 GW }  &
\textbf{3 GW }  &
\textbf{4 GW }  &\\
\midrule

O4 &
\cite{2021ApJ...909L..19G} &
20.5 &
20.6 &
19.5 &
--- &
\\
\\

 &
\cite{2019ApJ...876...18F} &
20.5 &
20.6 &
19.5 &
--- &
\\
\\

O5 &
\cite{2021ApJ...909L..19G} &
21.9 &
21.9 &
20.4 &
20.8 &
\\
\\

 &
\cite{2019ApJ...876...18F} &
21.9 &
21.9 &
20.3 &
20.8 &\\
\bottomrule
\end{tabular}
\caption{Summary of mean apparent AB magnitudes of discoverable KNe for LVK O4 and O5 based on simulations with 1, 2, 3, or 4 coincidental GW detection(s) and an EM detection.  The missing statistics for the 4-detector events during LVK O4 is due to a negligible number of mergers having coincidental detections on 4 instruments.}
\label{table:final-mags}
\end{table}

\subsection{LVK O4 observing run}\label{sec:LVKO4}

For this simulation we use both the \cite{2021ApJ...909L..19G} and \cite{2019ApJ...876...18F} mass distribution, a detection threshold of 23 magnitude, the DECam r passband for detections, the LVK User guide rates for BNS mergers, and a sun loss fraction of 0.5. In reality, the sun loss fraction is dependent on the specific follow-up survey we consider, the site(s) for ground-based facilities, and the location of the KN on the sky (and in particular, for space-based facilities the overlap between the GW localization and the allowed viewing area). Determining if an event will be lost to light from the sun must be done on an event by event basis, taking into account the survey strategy. Additionally, given that the search for real KNe will be done by a network of both public and private telescopes, which may elect to not share information about a counterpart for several hours after discovery, makes modeling this effect infeasible without several assumptions. For the sake of simplicity, we choose to encode this information using a constant $0.5$ fraction.

\begin{figure*}
\includegraphics[width=\linewidth]{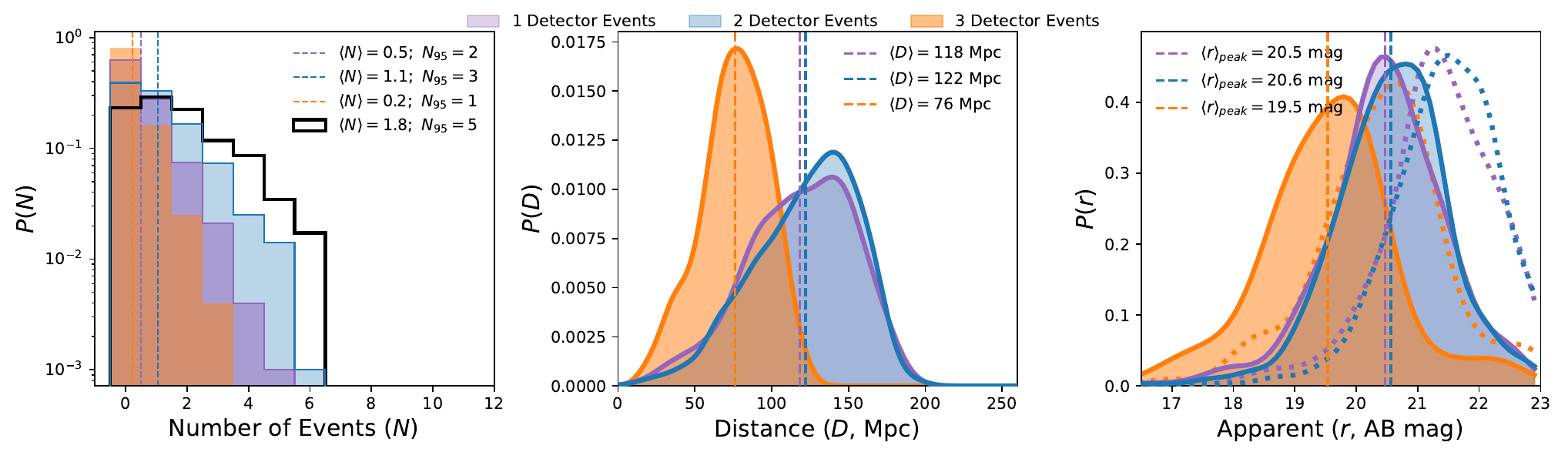}
\includegraphics[width=\linewidth]{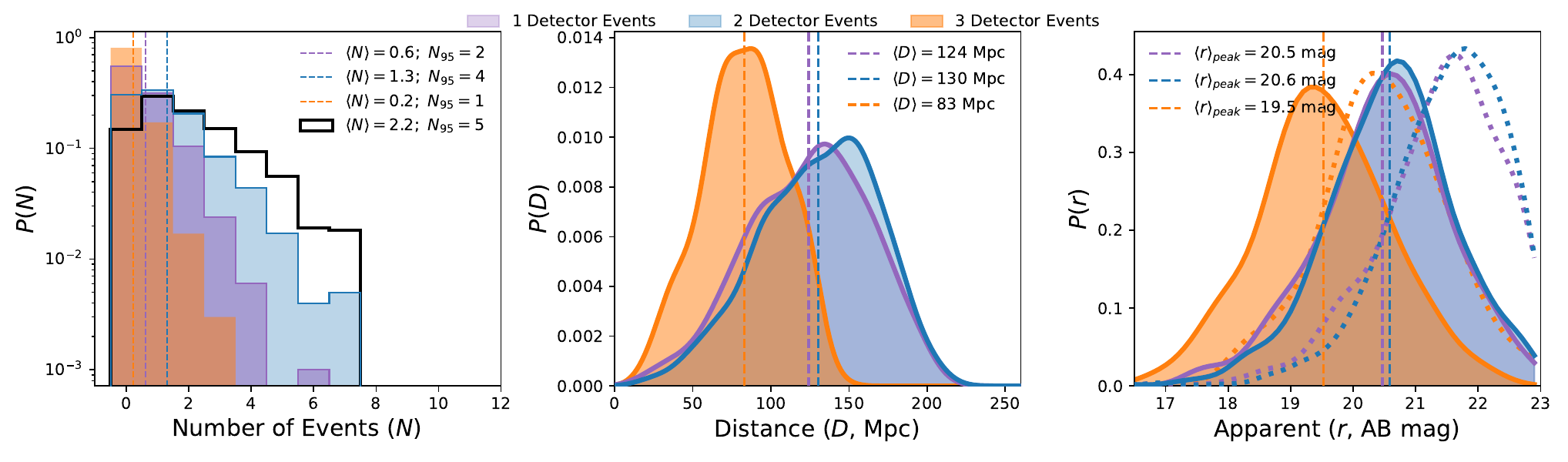}
    \caption{\textbf{LVK O4} Top pane shows results from using the \protect\cite{2019ApJ...876...18F} mass model while the bottom pane shows results from using the \protect\cite{2021ApJ...909L..19G} mass model.  \textbf{Left:} Distribution of the number of EM detectable events for 1, 2, and 3 GW detector events. Solid black line shows the distribution of the total number of discoverable kilonovae.\textbf{Center:} Distribution of distances of EM detectable events for 1, 2, and 3 GW detector events. \textbf{Right:} Distribution of peak and discovery magnitudes of EM detectable events for 1, 2, and 3 GW detector events. Solid lines represent distributions of peak magnitude, while dotted lines represent distributions of discovery magnitude. Table \ref{table:final-summary}, Table \ref{table:final-dist}, and Table \ref{table:final-mags} summarizes the results.} 
    \label{fig:mc-results-O4}
\end{figure*}

As evident from Table \ref{table:final-summary} and Figure \ref{fig:mc-results-O4}, the median number of mergers with detectable electromagnetic counterpart over all of the LVK O4 is $\sim 1 - 2$, depending on the mass model used. The figure also shows the expected values and distributions for event distances and magnitudes while Table \ref{table:final-summary} breaks down the events by the number coincidental GW detections.

Additionally, we found the distribution for the discovery window, defined as the time for which the kilonova is brighter than the limiting magnitude of the survey, for 1, 2, and 3-detector events (Table \ref{table:discovery-window}). Even though the times shown are shorter than the 10+ days of observations obtained for GW170817 (Figure \ref{fig:GW170817LC}), we should still, on average, have several days to get detections of the EM counterpart.

\begin{table}
\textbf{Discovery windows (days)}\\ \\
\begin{tabular}{lccc}

\toprule
\textbf{BNS mass model} &
\textbf{N} & 
 \textbf{90\% credible} &
\textbf{Mean}
\\
\midrule
\cite{2021ApJ...909L..19G} &
1 &  $3.4_{-2.0}^{+3.4}$  &  3.60\\ \\
 &
2 &  $3.2_{-2.0}^{+3.4}$  & 3.45\\ \\
 &
3  &  $4.4_{-2.2}^{+3.8}$   & 4.82\\ \\

\cite{2019ApJ...876...18F} &
1 &  $4.0_{-2.0}^{+3.0}$  &  4.29\\ \\
 &
2 &  $4.0_{-2.2}^{+3.0}$  & 4.13\\ \\
 &
3  &  $5.8_{-3.4}^{+3.2}$  & 5.82\\
\bottomrule
\end{tabular}
\caption{Discovery windows (in days) for KNe during LVK O4 for N = 1, 2, and 3  detector events in DECam  r-band. The missing statistics for the 4-detector events is due to a negligible number of mergers having coincidental detections on 4 instruments. All these discovery windows are significantly shorter than the 10+ days for which SSS17a was discoverable.}
\label{table:discovery-window}
\end{table}

\subsection{Looking ahead - LVK O5}

LVK O5 presents the next opportunity for finding kilonovae, post-GW trigger. With the observing run slated to begin at the end of 2026 with a proposed end in the middle of 2029, we expect the Vera Rubin Observatory to be operational for the entirety of O5. This section aims to paint a picture of what the next $\sim$ 5 years of kilonova discovery could look like. Once again, we use the most updated PSDs for the O5 run from LVK which, notably, represent the high end targets of BNS ranges for LIGO and KAGRA and the low end for Virgo. 

The predictions presented in this section must be assessed with the added context that the sensitivities used are the targeted, optimistic values and real PSDs during O5 might not achieve these goals. For this reason, another analysis for LVK O5 will likely be required once the PSDs and observing plans are defined more concretely. Regardless, we present tentative numbers here since they will be useful for planning and forecasting. 

\begin{figure*}
    \includegraphics[width=\linewidth]{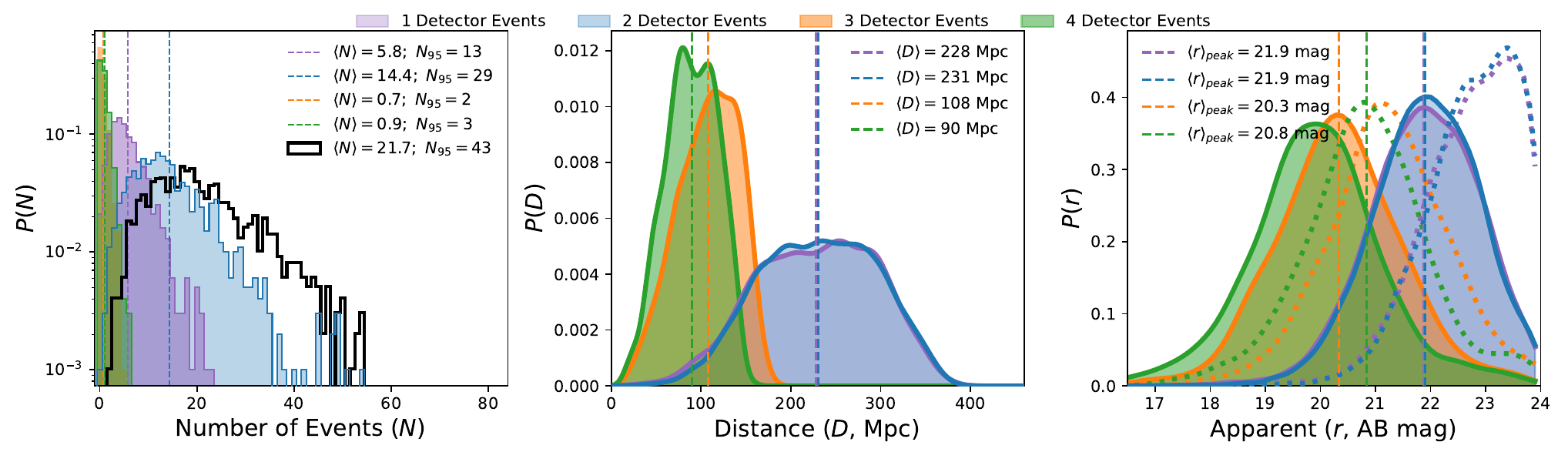}
    \includegraphics[width=\linewidth]{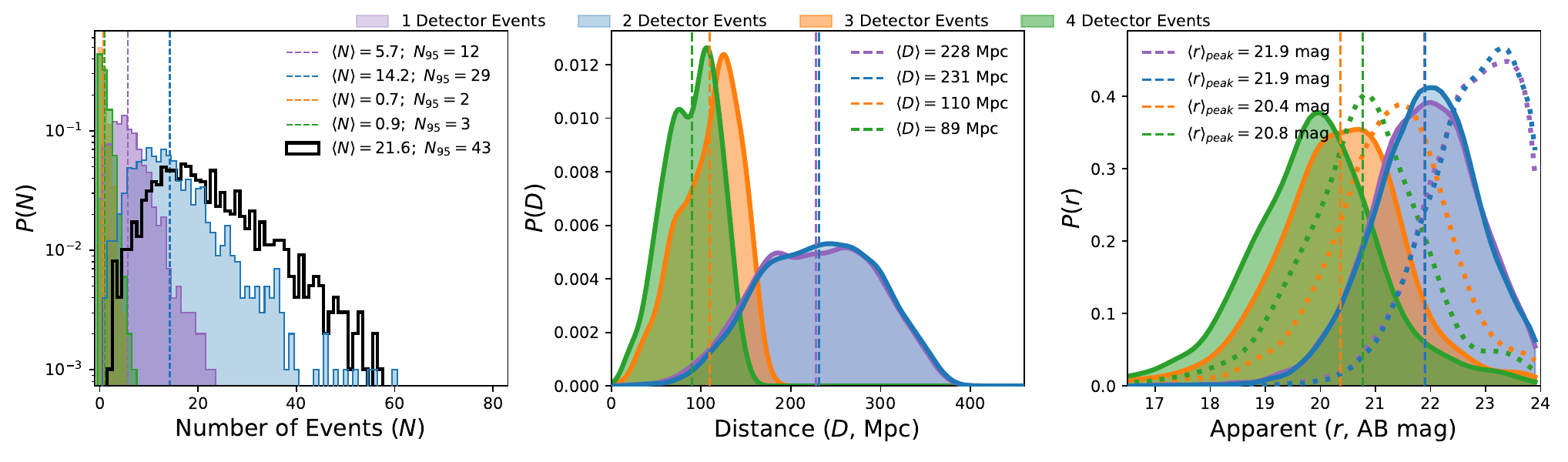}
    \caption{\textbf{LVK O5} Top pane shows results from using the \protect\cite{2019ApJ...876...18F} mass model while the bottom pane shows results from using the \protect\cite{2021ApJ...909L..19G} mass model.  \textbf{Left:} Distribution of the number of EM detectable events for 1, 2, 3, and 4 GW detector events. Solid black line shows the distribution of the total number of discoverable kilonovae. \textbf{Center:} Distribution of distances of EM detectable events for 1, 2, 3, and 4 GW detector events. \textbf{Right:} Distribution of peak and discovery magnitudes of EM detectable events for 1, 2, 3, and 4 GW detector events.Solid lines represent distributions of peak magnitude, while dotted lines represent distributions of discovery magnitude. Table \ref{table:final-summary}, Table \ref{table:final-dist}, and Table \ref{table:final-mags} summarizes the results.} 
    \label{fig:mc-results-O5}
\end{figure*}

With this caveat, we predict the median number of mergers with detectable electromagnetic counterpart over all of the LVK O5 to be $\sim 19$. As evident from Figure \ref{fig:mc-results-O5} and Table \ref{table:final-summary}, an updated LVK network during O5 with significantly improved sensitivities  may present the first opportunity to discover a small sample of kilonovae which would enable exciting new population studies furthering both transient astronomy and cosmology. 

\subsection{Comparison with current LVK O4 results}

To test the consistency of our pipeline with empirical observations, we simulate a partial LVK O4 run to compare with the actual ongoing LVK O4 run. We adjust the simulation parameters to match the current O4 run by setting:
\begin{itemize}
    \item the duration to be $\sim 4.67$ months, reflecting the current O4 period of May 24, 2023 to October 14, 2023;
    \item the duty cycles for Virgo and KAGRA to zero, since they are not observing during the current period;
    \item the duty cycles of the two LIGO detectors to $70\%$ $^{\ref{url:user-guide}}$.
\end{itemize}

All other configurable parameters mirror the full O4 simulation discussed in section \ref{sec:LVKO4}. 

This procedure allows us to assess the validity of our model, given that we have not detected any BNS mergers during the first $\sim 4.67$ months of LVK O4 (as of October 14, 2023). Using the \cite{2021ApJ...909L..19G} mass model, we found that the median number of disoverable kilonovae in the first $\sim4.67$ months of LVK O4, to be $0_{-0}^{+2}$ with $1_{-1}^{+3}$ BNS merger detections. 

\subsection{Comparison with complimentary work}

Complimentary work has been done in the past to understand the detection rates of KN for surveys like the Zwicky Transient Facility (ZTF), the Wide-Field Infrared Transient Explorer (WINTER), and the Vera Rubin Observatory \citep{2022ApJ...937...79C, 2022ApJ...926..152F, 2023arXiv230609234W}. Table \ref{table:KN-work-comparison} summarizes these results while Table \ref{table:final-summary} reports the results from this work. Our independent analysis with the ZTF - r band, a limiting magnitude of $21.4$, and the \cite{2021ApJ...909L..19G} mass model predicts the number of discoverable kilonovae to be ${ 1 }_{- 1 }^{+ 3}$ over the 18 month LVK O4.

\begin{table}
\begin{tabular}{lll}

\toprule
\textbf{Work} & \textbf{Band} & \textbf{KN Detections} \\
\midrule

\cite{2022ApJ...926..152F} &  J &  $1_{-1}^{+2}$ (over LVK O4) \\ \\
\cite{2023arXiv230609234W} &  r &  $0.43_{-0.26}^{+0.58}$ (per year) \\ \\
\cite{2022ApJ...937...79C} &  J &  $2.4_{-1.8}^{+3.6}$ (per year)\\ \\
\cite{2022ApJ...937...79C} &  r &  $5.1_{-3.8}^{+7.8}$ (per year) \\ 
\bottomrule
\end{tabular}
\caption{ Comparison of results from analysis for kilonova detection rates
from complimentary work for LVK O4.}
\label{table:KN-work-comparison}
\end{table}

\begin{figure}
    \includegraphics[width=\columnwidth]{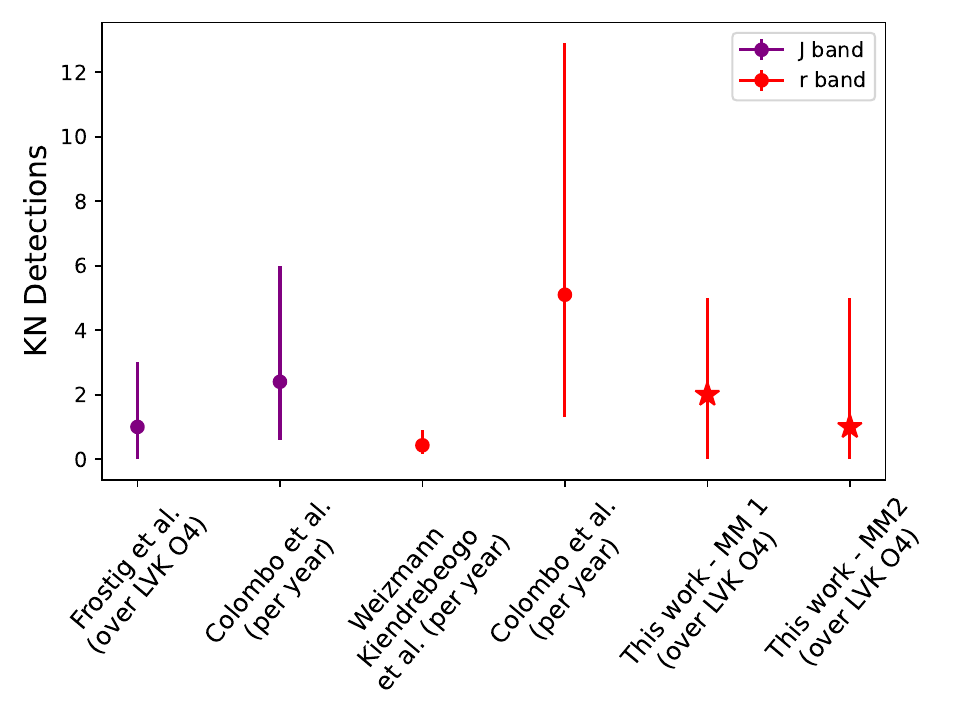}
    \caption{Comparison of our results to different KNe detection rate analysis done in the past for LVK O4. MM1 and MM2 correspond to the \protect\cite{2021ApJ...909L..19G} and \protect\cite{2019ApJ...876...18F} BNS mass models respectively} 
    \label{fig:kn_mags}
\end{figure}

\subsection{Retrospective analysis for LVC O2 and O3}

Since our flexible framework can easily adopt different PSDs over arbitrary observing durations to simulate discovery rates, we have performed our analysis for the LVC O2 and O3 observing campaigns. 
We note that this since the BNS merger rates used in this work are themselves inferred from the results of the O2 and O3 runs, these simulations are somewhat self fulfilling.  However, to the extent that this model accurately represent the true BNS merger rate, we can still gain valuable insights into the predictive capabilities of our framework. 

For this analysis, we use the PSDs from the GWTC-1 \citep{2019PhRvX...9c1040A}\footnote{https://dcc.ligo.org/LIGO-P1800374/public} and GWTC-2 catalogs \citep{2021PhRvX..11b1053A}\footnote{https://dcc.ligo.org/LIGO-P2000251/public}. The O2 campaign ran from November 30, 2016 to August 25, 2017 while the O3 campaign ran from April 1, 2019 to March 27, 2020. With the exception of the campaign duration, PSDs, and the absence of KAGRA, all other simulation configurations were identical to the ones used for LVK O4. Additionally, we exclusively used the \cite{2021ApJ...909L..19G} BNS mass model for this exercise. 

We predict the number of detectable BNS mergers over the O2 run to be ${1}_{-1}^{+2}$ with ${0}_{-0}^{+2}$ discoverable KNe (90\% credible). Only $\sim$32\% of our trials had $ \geq 1$ discoverable KNe during this period, consistent with the single discovery of GW170817 \citep{LIGOScientific:2017vwq} and SSS17a \citep{2017Sci...358.1556C} during O2.

For the O3 run, we predict the number of detectable BNS mergers to be ${1}_{-1}^{+3}$ with ${0}_{-0 }^{+2}$ discoverable KNe (90\% credible). Despite the increased sensitivity from O2, only $\sim$49\% of our trials had $\geq$1 discoverable KNe during this period.  Nevertheless, these results are also consistent with the single discovery of GW190425 \citep{2020ApJ...892L...3A} and no KN discovery during O3.

\section{Conclusion}

These results paint kilonovae like SSS2017a with confirmed GW detections, gamma ray bursts, counterpart photometry, and spectroscopy as incredibly rare. The low intrinsic rate of detectable kilonovae is compounded by difficulties in locating the counterparts in poorly localized skymaps that, at this time, routinely have $90\%$ confidence intervals that span on the order of $10^3 \text{sq deg}$. Such large localization are simply impractical to probe efficiently without wide field of view surveys \citep{2023arXiv230609234W}. These inefficiencies are difficult to model accurately and thus the numbers presented here are upper limits.

Moreover, Table \ref{table:discovery-window} illustrates how our window of opportunity for finding future kilonovae will likely be significantly shorter than GW170817's counterpart. These factors demonstrate the need for improved tooling, infrastructure, and search strategies for future KN discovery, such as \textsc{Teglon} \footnote{https://github.com/davecoulter/teglon} and the systems described in \cite{2021MNRAS.504.2822A}, \cite{2023arXiv230204878B} and \cite{2022MNRAS.509..914C}.

Finally, a prompt chirp mass estimate from LVK, even if provided only to a tenth of a solar mass or with a small random offset applied, would allow forecasting of the electromagnetic signal. This, in turn, would enable observers to prioritize and coordinate follow-up resources more effectively, improving the yield of counterpart discoveries. Our synthetic photometry pipeline can be integrated into alerts systems, like the one described in Section \ref{sec:slacbot}, to inform discovery and follow up strategies.

\section{Data Availability}

\textit{\textbf{Software: }} This work makes use of \textsc{Numpy} \citep{numpy}, \textsc{Astropy} \citep{astropy:2013, astropy:2018, astropy:2022}, \textsc{Sncosmo} \citep{sncosmo}, \textsc{Kilopop} \citep{Setzer2023}, \textsc{Scipy} \citep{2020SciPy-NMeth}, \textsc{Ligo em bright} \footnote{https://git.ligo.org/emfollow/em-properties/em-bright}, \textsc{Inspiral range} \footnote{https://git.ligo.org/gwinc/inspiral-range}, 
\textsc{Possis} \footnote{https://github.com/mbulla/kilonova\_models}, \textsc{Matplotlib} \citep{matplotlib}, and \textsc{Pandas} \citep{pandas1, pandas2}.
\\
\\
\textit{\textbf{Data availability:}} All the code used in this work is publicly available at \url{https://github.com/uiucsn/KNmodel}. 

\section{Acknowledgments}

VS acknowledges the support of the LSST Corporation's 2021 Enabling Science award for undergraduates for making this work possible, as well as travel support to present this research at the LSST Project and Community Workshop in 2023. GN gratefully acknowledges NSF support from  AST-2206195, and a CAREER grant, supported in-part by funding from Charles Simonyi. HP is supported by an Illinois Center for AstroPhysical Surveys Graduate Student Fellowship. DC would like to acknowledge support from the NSF grants OAC-2117997 and PHY-1764464. BC is supported by the NSF Graduate Research Fellowship Program under Grant No. DGE 21-46756.

The UCSC team is supported in part by NASA grant NNG17PX03C, NSF grants AST-1815935 and AST-2307710, the Gordon \& Betty Moore Foundation, and by a fellowship from the David and Lucile Packard Foundation to R.J.F.

The \textsc{Slack} application developed by this work is hosted by the Scalable Cyberinfrastructure for Multi-Messenger Astrophysics group (SCiMMA, \url{https://scimma.org}, PI: Narayan), which is supported by the National Science Foundation through the Office of Advanced Cyberinfrastructure awards OAC-1841625, OAC-1934752, and OAC-2311355. It is also deployed on the ANTARES broker system (\url{https://antares.noirlab.edu}) hosted by NSF's National Optical and Infrared Research Laboratory (NOIRLab). This work made use of the Illinois Campus Cluster, a computing resource that is operated by the Illinois Campus Cluster Program (ICCP) in conjunction with the National Center for Supercomputing Applications (NCSA) and which is supported by funds from the University of Illinois at Urbana-Champaign. This work was partially supported by the Center for AstroPhysical Surveys (CAPS) at the National Center for Supercomputing Applications (NCSA), University of Illinois Urbana-Champaign. This work makes extensive use of software and public data products produced by the LIGO Scientific Collaboration. The work of the LSC is supported by NSF’s LIGO Laboratory which is a major facility fully funded by the National Science Foundation.

\bibliographystyle{mnras}
\bibliography{citations} 
\clearpage
\appendix 

\section{Slack bot}
\label{sec:slacbot}

\begin{figure}
    \includegraphics[width=\columnwidth]{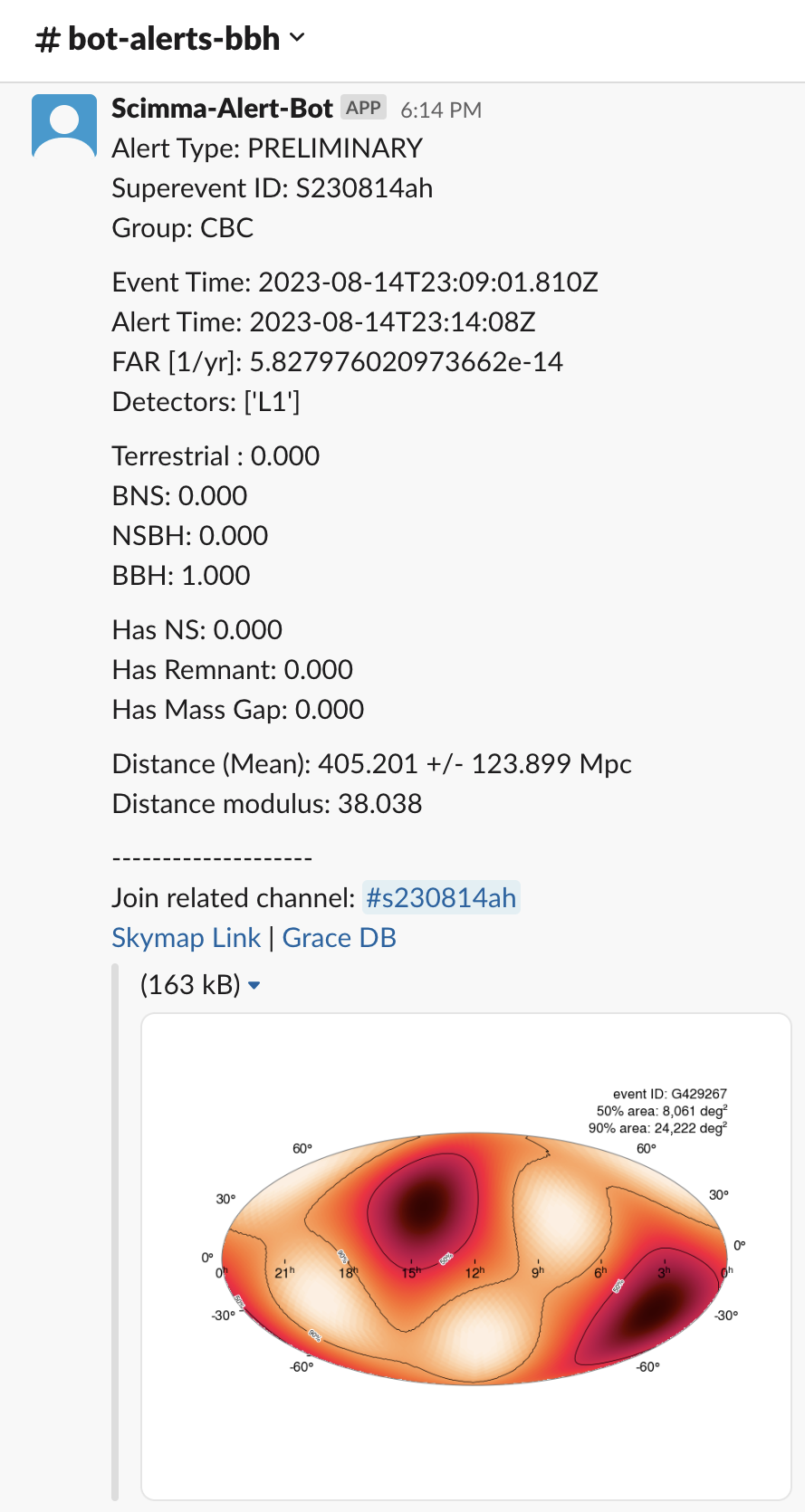}
    \caption{Screen capture of LVK alert streamed to \textsc{Slack} workspace} 
    \label{fig:slackbot}
\end{figure}

In order to facilitate future kilonovae discovery, we created a bot that streams LVK compact binary coalescence (CBC) and burst alerts to \textsc{Slack} workspaces (Figure \ref{fig:slackbot}) using Scimma's Hopskotch \footnote{https://scimma.org/hopskotch.html}. This bot can be configured to filter alerts by the false alarm rate, likelihood of being a BNS or NSBH merger, having a luminous remnant, distance etc. It can also create different channels for events to facilitate event specific discussion. The bot is open source and publicly available at \url{https://github.com/scimma/slackbot} and is already operational on The Gravity Collective \citep{2021ApJ...923..258K} and ANTARES \citep{2021AJ....161..107M}\footnote{https://antares.noirlab.edu/loci} workspaces.

\section{Photometry credit}
\label{photometry-credit}

Figure \ref{fig:GW170817LC} uses photometry that was originally collected by \cite{2017PASA...34...69A}, \cite{2017Natur.551...64A}, 
\cite{2017Sci...358.1556C}, 
\cite{2017ApJ...848L..17C},
\cite{2017ApJ...848L..29D},
\cite{2017Sci...358.1570D},
\cite{2017Sci...358.1565E},
\cite{2017SciBu..62.1433H},
\cite{2017Sci...358.1559K},
\cite{2017ApJ...850L...1L},
\cite{2017Natur.551...67P},
\cite{2018ApJ...852L..30P},
\cite{2017Sci...358.1574S},
\cite{2017Natur.551...75S},
\cite{2017ApJ...848L..27T},
\cite{2017Natur.551...71T},
\cite{2017PASJ...69..101U}, and
\cite{2017ApJ...848L..24V}.
As requested in the paper with the combined data, please ensure that any use of this photometry includes appropriate citation to the original papers, in addition to the paper that compiled all the data \citep{Villar_2017}.

\section{BNS Horizon Distances}

In order to compute the BNS horizon distances (HD) we use the instrument PSD files mentioned in Table \ref{table:psd}.Table \ref{table:hd} show the minimum $(\mathrm{m_1=1M_{\odot}, m_2=1M_{\odot}})$ and maximum  $\mathrm{(m_1=2.05M_{\odot}, m_2=2.05M_{\odot})}$ BNS horizon distances for the LVK O4 and O5 observing runs. These were also used to determine the dimensions of the box in which the mergers would take place.

\begin{table}
\begin{tabular}{lll}

\toprule
\textbf{Instrument} & \textbf{O4 PSD File} & \textbf{O5 PSD File}   \\
\midrule
LIGO - Livingston & aligo\_O4high.txt & AplusDesign.txt \\
LIGO - Hanford  & aligo\_O4high.txt  &  AplusDesign.txt \\ 
Virgo &  avirgo\_O4high\_NEW.txt & avirgo\_O5low\_NEW.txt  \\
KAGRA & KAGRA\_10Mpc.txt  & kagra\_128Mpc.txt   \\

\bottomrule
\end{tabular}
\caption{PSDs used for LVK O4 and O5 simulation work.}
\label{table:psd}
\end{table}

\begin{table}
\begin{tabular}{llll}

\toprule
\textbf{Observing Run} & \textbf{Instrument} & \textbf{Min HD (MPc)}  & \textbf{Max HD (MPc)} \\
\midrule
O4 & LIGO & 140.68 & 252.23\\
   & Virgo & 90.13 & 162.41\\
   & KAGRA & 7.78 & 14.13\\
O5 & LIGO & 253.39 & 449.47\\
   & Virgo & 113.00 & 203.16\\
   & KAGRA & 100.12 & 180.22\\
\bottomrule
\end{tabular}
\caption{Minimum and maximum horizon distances for LVK O4 and O5 observing runs.}
\label{table:hd}
\end{table}

\section{SED Scaling Parameters}

This section documents the scaling parameters used for the SED extrapolation process described in Section \ref{sed}.

\begin{table}
\begin{tabular}{lll}

\toprule
\textbf{Parameter} & \textbf{Name} & \textbf{Sum of residuals}  \\
\midrule
$c$ & Intercept &  $0.0043$  \\
$m$ & Slope &  $9.5348$  \\
$n$ & Exponent & $8.2920e-05$ \\
\bottomrule
\end{tabular}
\caption{Sum of residuals for the spline surfaces created for all three scaling parameters}
\label{table:residuals}
\end{table}

\begin{figure*}
    \includegraphics[width=0.99\linewidth]{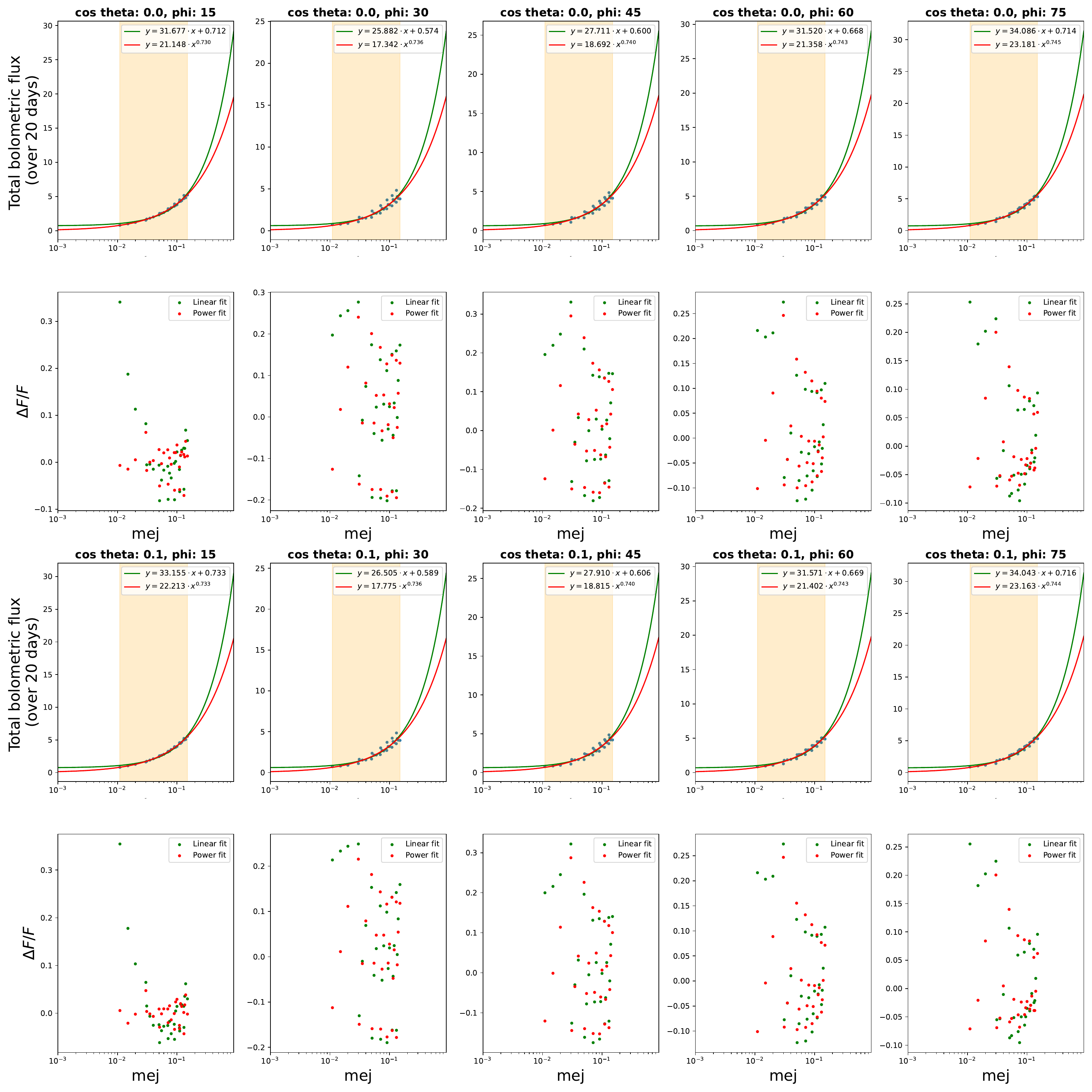}
    \caption{Scaling laws for all $\Phi$ and $\textrm{cos }\Theta = 0 \textrm{ or 0.1}$ pairs along with the $\mathrm{\Delta F/F}$ errors. All parameters for the linear scaling laws and the power scaling laws are provided in table \ref{table:linear-laws} and \ref{table:power-laws} respectively.} 
    \label{fig:scaling-laws}
\end{figure*}

\begin{table}
\centering
\begin{tabular}{llll}
\toprule
 cos\_theta & phi &     slope &  intercept \\
\midrule
       0.0 &  15 & 31.676920 &   0.711811 \\
       0.0 &  30 & 25.882260 &   0.574318 \\
       0.0 &  45 & 27.711210 &   0.600171 \\
       0.0 &  60 & 31.520312 &   0.668417 \\
       0.0 &  75 & 34.086420 &   0.714488 \\
       0.1 &  15 & 33.154611 &   0.732685 \\
       0.1 &  30 & 26.505415 &   0.588698 \\
       0.1 &  45 & 27.910461 &   0.606310 \\
       0.1 &  60 & 31.571041 &   0.668550 \\
       0.1 &  75 & 34.043173 &   0.715819 \\
       0.2 &  15 & 36.546367 &   0.800088 \\
       0.2 &  30 & 28.526260 &   0.634880 \\
       0.2 &  45 & 28.587370 &   0.626385 \\
       0.2 &  60 & 31.732794 &   0.674077 \\
       0.2 &  75 & 34.009856 &   0.715993 \\
       0.3 &  15 & 39.916188 &   0.868458 \\
       0.3 &  30 & 32.279970 &   0.715132 \\
       0.3 &  45 & 30.015271 &   0.663718 \\
       0.3 &  60 & 32.140097 &   0.685910 \\
       0.3 &  75 & 34.059853 &   0.716032 \\
       0.4 &  15 & 43.143247 &   0.931342 \\
       0.4 &  30 & 37.554645 &   0.822008 \\
       0.4 &  45 & 32.495046 &   0.719034 \\
       0.4 &  60 & 32.918693 &   0.703164 \\
       0.4 &  75 & 34.180384 &   0.717938 \\
       0.5 &  15 & 46.681507 &   0.999113 \\
       0.5 &  30 & 43.192080 &   0.933819 \\
       0.5 &  45 & 36.254005 &   0.798361 \\
       0.5 &  60 & 34.184250 &   0.731180 \\
       0.5 &  75 & 34.385484 &   0.721941 \\
       0.6 &  15 & 50.388877 &   1.070547 \\
       0.6 &  30 & 48.115833 &   1.026022 \\
       0.6 &  45 & 41.308726 &   0.905888 \\
       0.6 &  60 & 36.168610 &   0.775022 \\
       0.6 &  75 & 34.729490 &   0.728454 \\
       0.7 &  15 & 54.182222 &   1.144727 \\
       0.7 &  30 & 53.329365 &   1.121405 \\
       0.7 &  45 & 47.011114 &   1.029205 \\
       0.7 &  60 & 39.336503 &   0.837144 \\
       0.7 &  75 & 35.277649 &   0.738860 \\
       0.8 &  15 & 58.138435 &   1.215019 \\
       0.8 &  30 & 58.757791 &   1.219488 \\
       0.8 &  45 & 52.297437 &   1.127138 \\
       0.8 &  60 & 43.892658 &   0.936113 \\
       0.8 &  75 & 36.259997 &   0.755234 \\
       0.9 &  15 & 62.179061 &   1.283004 \\
       0.9 &  30 & 64.414780 &   1.315016 \\
       0.9 &  45 & 58.136458 &   1.228854 \\
       0.9 &  60 & 48.871449 &   1.035869 \\
       0.9 &  75 & 38.187313 &   0.793950 \\
       1.0 &  15 & 67.452758 &   1.331387 \\
       1.0 &  30 & 71.137258 &   1.398322 \\
       1.0 &  45 & 65.139861 &   1.323359 \\
       1.0 &  60 & 54.699193 &   1.110350 \\
       1.0 &  75 & 41.802789 &   0.863606 \\
\bottomrule
\end{tabular}
\caption{Parameters for linear scaling }
\label{table:linear-laws}
\end{table}

\begin{table}
\centering
\label{table:power-laws}
\begin{tabular}{llll}
\toprule
 cos\_theta & phi &  coefficient &  exponent \\
\midrule
       0.0 &  15 &    21.147741 &  0.730248 \\
       0.0 &  30 &    17.341877 &  0.736150 \\
       0.0 &  45 &    18.692025 &  0.740490 \\
       0.0 &  60 &    21.357779 &  0.742585 \\
       0.0 &  75 &    23.180504 &  0.744509 \\
       0.1 &  15 &    22.212722 &  0.733037 \\
       0.1 &  30 &    17.774688 &  0.735778 \\
       0.1 &  45 &    18.815112 &  0.739729 \\
       0.1 &  60 &    21.402487 &  0.742854 \\
       0.1 &  75 &    23.162582 &  0.744420 \\
       0.2 &  15 &    24.495838 &  0.734150 \\
       0.2 &  30 &    19.144498 &  0.734297 \\
       0.2 &  45 &    19.232806 &  0.737476 \\
       0.2 &  60 &    21.488825 &  0.742021 \\
       0.2 &  75 &    23.121571 &  0.743968 \\
       0.3 &  15 &    26.746196 &  0.735150 \\
       0.3 &  30 &    21.720584 &  0.734577 \\
       0.3 &  45 &    20.162467 &  0.735113 \\
       0.3 &  60 &    21.747593 &  0.741105 \\
       0.3 &  75 &    23.139055 &  0.743795 \\
       0.4 &  15 &    28.823373 &  0.735425 \\
       0.4 &  30 &    25.218208 &  0.735069 \\
       0.4 &  45 &    21.794859 &  0.733628 \\
       0.4 &  60 &    22.239784 &  0.740152 \\
       0.4 &  75 &    23.207371 &  0.743607 \\
       0.5 &  15 &    31.089173 &  0.735924 \\
       0.5 &  30 &    28.803779 &  0.734589 \\
       0.5 &  45 &    24.244122 &  0.732731 \\
       0.5 &  60 &    23.046518 &  0.738911 \\
       0.5 &  75 &    23.334746 &  0.743398 \\
       0.6 &  15 &    33.400762 &  0.735875 \\
       0.6 &  30 &    31.937504 &  0.735144 \\
       0.6 &  45 &    27.451223 &  0.731421 \\
       0.6 &  60 &    24.333332 &  0.737719 \\
       0.6 &  75 &    23.555430 &  0.743193 \\
       0.7 &  15 &    35.777313 &  0.736108 \\
       0.7 &  30 &    35.236191 &  0.735871 \\
       0.7 &  45 &    31.016546 &  0.730061 \\
       0.7 &  60 &    26.383905 &  0.737269 \\
       0.7 &  75 &    23.906473 &  0.742878 \\
       0.8 &  15 &    38.294384 &  0.737310 \\
       0.8 &  30 &    38.685709 &  0.736969 \\
       0.8 &  45 &    34.413624 &  0.731455 \\
       0.8 &  60 &    29.287982 &  0.735747 \\
       0.8 &  75 &    24.538580 &  0.742689 \\
       0.9 &  15 &    40.957331 &  0.739484 \\
       0.9 &  30 &    42.340656 &  0.739015 \\
       0.9 &  45 &    38.211447 &  0.733714 \\
       0.9 &  60 &    32.539923 &  0.736270 \\
       0.9 &  75 &    25.795598 &  0.741992 \\
       1.0 &  15 &    44.993768 &  0.748327 \\
       1.0 &  30 &    47.201948 &  0.746468 \\
       1.0 &  45 &    43.143204 &  0.740656 \\
       1.0 &  60 &    36.690299 &  0.743248 \\
       1.0 &  75 &    28.286827 &  0.743213 \\
\bottomrule
\end{tabular}
\caption{Parameters for power scaling }
\label{table:power-laws}
\end{table}

\end{document}